\documentclass[11pt]{article}

\usepackage{amsfonts}

\usepackage{amsmath}
\usepackage{amssymb}
\usepackage{latexsym}
\usepackage{graphicx}
\usepackage[english]{babel}
\usepackage[font={small}]{caption}

\usepackage{amsfonts}
\usepackage{latexsym}
\usepackage{graphicx}
\usepackage[english]{babel}
\usepackage{tikz}

\begin{document}
\input epsf

\def\p{\partial}
\def\h{{1\over 2}}
\def\be{\begin{equation}}
\def\bea{\begin{eqnarray}}
\def\ee{\end{equation}}
\def\eea{\end{eqnarray}}
\def\d{\partial}
\def\la{\lambda}
\def\eps{\epsilon}
\def\b{\bigskip}
\def\m{\medskip}

\newcommand{\newsection}[1]{\section{#1} \setcounter{equation}{0}}

\def\q{\quad}

\def\h{{1\over 2}}
\def\t{\tilde}
\def\r{\rightarrow}
\def\nn{\nonumber\\}

\let\p=\partial

\begin{flushright}
\end{flushright}
\vspace{20mm}
\begin{center}
{\LARGE Can we observe fuzzballs or firewalls?}
\\
\vspace{18mm}
Bin Guo\footnote{guo.1281@osu.edu}, Shaun Hampton\footnote{hampton.197@osu.edu} and   Samir D. Mathur\footnote{mathur.16@osu.edu}

\vskip .1 in

 Department of Physics\\The Ohio State University\\ Columbus,
OH 43210, USA\\
\vspace{4mm}

\end{center}
\vspace{10mm}
\thispagestyle{empty}
\begin{abstract}

In the fuzzball paradigm the information  paradox is resolved because the black hole is replaced  by an object with no horizon.  One may therefore ask if observations can distinguish a traditional hole from a fuzzball. We find: (a) It is very difficult to reflect quanta  off the surface of a fuzzball, mainly  because geodesics starting near the horizon radius cannot escape to infinity   unless their starting direction is very close to radial. (b) If infalling particles interact with the emerging radiation  before they are engulfed by the horizon, then we say that  we have a `firewall behavior'. We consider several types of interactions, but find no evidence for firewall behavior in any theory that obeys causality. (c) Photons with  wavelengths {\it larger} than the black hole radius can be scattered off the emerging radiation, but a very small fraction of the backscattered photons will be able to escape back to infinity. 

\end{abstract}
\vskip 1.0 true in

\newpage
\setcounter{page}{1}

\numberwithin{equation}{section} 

\section{Introduction}

In the traditional picture of a black hole, all the mass resides at a central singularity, while the rest of spacetime is empty. In particular, the region around the horizon is in a vacuum state. 

But such a picture of the hole leads to the black hole information paradox \cite{hawking}. In string theory, it appears that this paradox is avoided because the structure of the hole is radically different:  the hole is described by a horizon sized {\it fuzzball}, with no horizon \cite{lm4,fuzzballs}. In this paper we explore the possibility of observing signatures of the fuzzball (and other associated effects)  in astronomical observations. We will argue that while such signatures are possible in principle, a variety of factors will likely make them very difficult to detect. 

\subsection{The information paradox}

 Suppose we assume the traditional picture of the hole, which has smooth spacetime in the vicinity of the horizon. The vacuum around such a horizon is unstable to the creation of particle-antiparticle pairs. One member of the pair escapes to infinity as Hawking radiation, while the other member (carrying negative energy)  falls into the hole to reduce its mass. The two members of the pair are in an entangled state, so the entanglement of the radiation with the remaining hole keeps growing as the evaporation proceeds. If the hole evaporates away completely we are left with radiation that is entangled, but there is nothing that it is entangled {\it with}. Such a configuration cannot be described by a pure state $\psi_{radiation}$, but only be a density matrix $\rho_{radiation}$. Thus the initial pure state of a collapsing star $\psi_{star}$ has evolved not to a new pure state but to a mixed state, in violation of the basic evolution in quantum mechanics. This led Hawking to argue in 1975 that the formation and evaporation of black holes violates the unitarity of quantum theory \cite{hawking}. 

It has sometimes been suggested that Hawking's computation of entanglement was a leading order computation, and small quantum gravity effects may introduce delicate correlations among the emitted quanta, so that the overall radiation state is actually {\it not} entangled with the hole near the endpoint of evaporation. In particular, Hawking's retraction in 2004 of his claim that black holes posed a problem was based on similar lines \cite{hawkingreverse}. He noted that there could be exponentially small corrections to the evaporation process due subleading saddle points in quantum gravity, and it was possible that these corrections would resolve the problem of unitarity. 

But such possibilities were removed by the  ``small corrections theorem" proved in 2009 \cite{cern}. Suppose the quantum gravity effects introduce a correction bounded by $\epsilon$ to the state of each created pair, with $\epsilon\ll 1$. Then using the strong subadditivity of quantum entanglement entropy, it can be proved that the reduction of entanglement $\delta S_{ent}$ due to these corrections is bounded as 
\be
{\delta S_{ent}\over S_{ent}}< 2\epsilon
\label{one}
\ee
Thus the Hawking argument of 1975  becomes a `theorem': unless we have O(1) corrections to low energy physics around the horizon, we cannot resolve the problem of growing entanglement between the radiation and the remaining hole. 

This leaves us with three choices:

\b

(i) {\it Remnants:}  The hole has the traditional horizon  to leading order while it is large, but perhaps the  evaporation stops due to quantum gravity effects when the hole gets to planck size.  Then the information in the initial star  and in  the negative energy particles that fell in gets locked up in an object with mass $\sim m_p$ and radius $\sim l_p$. This option is favored by many relativists; for example the remnant has been modeled as a baby universe. As we will note below however, string theory does not allow remnants, if we assume that AdS/CFT duality is correct. 

\m

(ii) {\it Fuzzballs:}  The state of the hole is  {\it not} given by the semiclassical state  up to small corrections. The microstates of the hole are `fuzzballs' with no horizon \cite{lm4,fuzzballs}. These microstates radiate from their surface just like a normal warm body (and not through a process of pair creation from the vacuum at a  horizon). This removes the information paradox. The nontriviality of the fuzzball construction is that the fuzzball resists the gravitational collapse that is a general feature of Einstein gravity coupled to normal (i.e. not stringy) matter.  

\m

(iii) {\it Nonlocality:}  We keep the traditional horizon with vacuum in its vicinity, but  postulate nonlocal effects in the quantum gravity theory that get around the information problem. Maldacena and Susskind have postulated that the radiation quanta near infinity are connected back to the interior of the hole through tiny wormholes \cite{cool}. Giddings has postulated nonlocality for low energy modes over distances of order the horizon scale \cite{giddings}. Hawking, Perry and Strominger have recently argued that information is stored even more nonlocally, in zero modes of the gauge parameter that extend to infinity \cite{hps}.

\b

In this paper we will be concerned with  the observability of fuzzballs. Also, it has been argued that objects like fuzzballs must exhibit a `firewall' behavior, where incoming objects get burnt by Hawking radiation before reaching the surface of such an object \cite{amps}. We will note that this postulate needs a modification \cite{flaw}, and then ask if this modified firewall behavior can be observed.

\subsection{Fuzzballs}

The information puzzle would have a  straightforward resolution if  the black hole was like a `planet with a surface'  rather than a `spacetime with the vacuum state around a horizon'. As we will note in more detail below, the obstruction to such a resolution has always been that it is very hard to make a horizon sized ball that will not undergo runaway collapse to a singularity. For example, the Buchdahl theorem  says that a perfect fluid ball with pressure decreasing outwards and with radius $R<{9\over 4} M$ cannot exist, since the pressure needed to support it against gravitational collapse would need to diverge to infinity somewhere in its interior \cite{buchdahl}. 

Remarkably, it was found that in string theory, black hole microstates are in fact horizon sized `fuzzballs' \cite{emission, lm4,  fuzzballs}. The novel features of the theory -- extended objects, extra dimensions, Chern-Simmons terms etc. -- bypass the traditional no-go theorems that required black holes to have  no `hair' \cite{gibbonswarner}. 

\begin{figure}[htbp]
   \begin{center}
   \includegraphics[scale=.8]{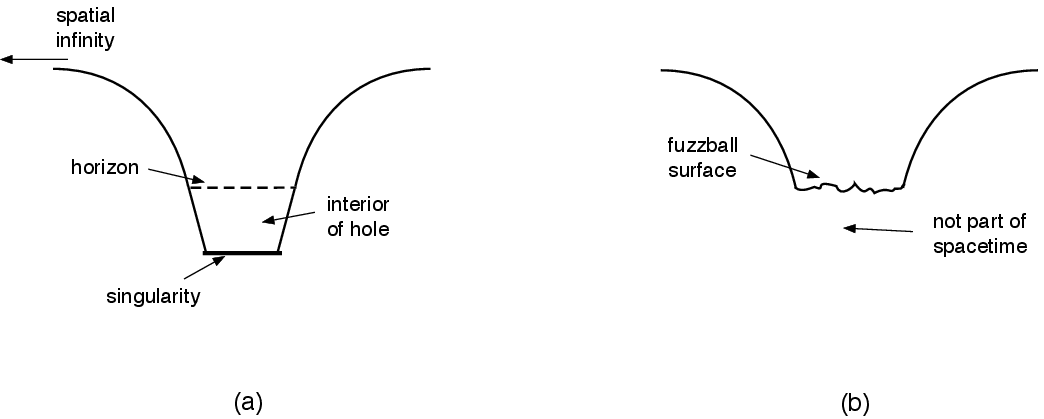}
   \caption{(a) The conventional picture of a black hole.  (b) the fuzzball picture: spacetime ends just outside the horizon in a quantum mess. }
   \label{ffuzzball}
   \end{center}
   \end{figure}

String theory has no free parameters: the brane content and interaction properties are all uniquely fixed by the requirements of quantum consistency. Black hole microstates must be made as bound states of the given objects in the theory. In \cite{emission} it was found that the size of such bound states grows with the number of branes in the bound state, in such  way that the size of the bound state is always order the horizon size for the given mass and charge. Next, brane bound states were studied one by one, starting with the simplest ones and moving to ones of more complexity. In all cases which have been examined over the years, the microstate was found to have no horizon. The fuzzball conjecture says that no microstate of any black holes in string theory will have a horizon;  i.e. there will be no region of spacetime which to leading order has the vacuum structure associated to a traditional picture of the horizon. This conjecture is illustrated pictorially in fig.\ref{ffuzzball}. 

Since the fuzzball has a surface rather than a vacuum region around a horizon, one can in principle probe the structure of this surface by observations. But it turns out that such observations are likely to be difficult, for a combination of reasons. While the generic fuzzball state has not been constructed for neutral holes, we will give arguments for why the boundary of such fuzzballs must lie at 
\be
r_b=2GM+\epsilon
\ee
with $\epsilon\ll GM$. Outside the radius $r_b$ the metric is expected to be very close to the traditional Schwarzschild metric. We should therefore  look at light rays in the Schwarzschild metric emanating from the location $r_b$. For $\epsilon\ll GM$, one finds that most of these rays fall back onto the surface at $r_b$; only rays that are very close to the radially outward direction manage to escape.\footnote{Similar observations have also appeared in a nice recent paper \cite{cardoso} which gives a detailed discussion of the observability of black holes.} For $\epsilon$ small, we therefore find a very small probability of detecting rays that bounce off the fuzzball. 

The difficulty of observation is further enhanced by the fact that the fuzzball surface is likely to be highly absorptive, so that it is not easy to bounce a light ray off it in the first place \cite{beyond}. The reason for this can be traced back to a fundamental property of the black hole: its very large entropy $S={A/4G}$. The energy gap between black hole microstates is  very small
\be
\Delta E \sim e^{-S}M
\label{bthree}
\ee
An infalling quantum with energy $E>\Delta E$ will not just explore the given microstate; it will instead cause a jump from the given microstate to a dense band of neighboring energy eigenstates.\footnote{In a recent work \cite{warnernew} this small energy gap was explicitly found in a family of fuzzball constructions.} There is much more phase space for the energy $E$ to reside in excitations of the fuzzball microstate, as compared to residing in a single reflected quantum of energy $E$. This makes the fuzzball surface highly absorptive, and it is very unlikely that a quantum that hits it will bounce back. In other words, the probability of reflection $P_{reflect}$ is expected to be tiny:
\be
P_{reflect} \ll 1
\label{preflect}
\ee

These factors make it hard to distinguish the fuzzball from a traditional black hole. Both objects  have the Schwarzschild metric at $r>2GM$ and both are highly absorptive for rays that reach close to the horizon. We will in fact argue that this similarity between fuzzballs and the traditional hole may be evidence for a general `correspondence principle', which states that the fuzzball mimics the traditional black hole for all but the most delicate quantum observations. 

\subsection{Firewall behavior}

In \cite{amps} it was argued that if we replace the traditional black hole with something that has a surface at $r=2GM+\epsilon$, then an infalling object will get burnt by the radiation emanating from this surface. If such a burning indeed happens, then we would say that the infalling object has encountered a `firewall'.   Note that  the firewall is not a {\it construction} of an object with a surface; rather it is a general argument along the following lines.   Suppose we replace the horizon with a hot surface just outside the horizon at $r_b=2GM+\epsilon$.  We assume that this surface radiates information unitarily just as a hot body does. Suppose we add one further assumption: that  there are no novel quantum gravity effects in the region $r>r_b$; i.e., physics outside the object is normal physics (`effective field theory'). The firewall argument then says the following: in this situation, an infalling object will interact with quanta of increasingly high energy as it approaches $r_b$, and the object will therefore get burnt by a firewall of such quanta. 

The intuition behind the firewall argument is simple. In Hawking's 1975 computation \cite{hawking}, radiation was created by pair production from the vacuum. In this process, the radiated modes do not get populated by real quanta (i.e., quanta that can be interacted with) until they are well away from the horizon (say, at $r\gtrsim 4GM$). But if the quanta are radiated from a hot surface at $r_b=2GM+\epsilon$, and they travel out as dictated by  normal physics for $r>r_b$, then an infalling object can interact with them at {\it any} location $r>r_b$. Because of the large redshift between $r_b$ and infinity, these quanta will be very energetic near $r_b$, reaching planck energies at $r_b$ if $\epsilon\sim {l_p^2 \over GM}$. 
An infalling object will interact with and get burnt by these high energy quanta before reaching the surface at $r_b$. In \cite{amps} a nice argument making this precise was given using the bit model used in the small corrections theorem \cite{cern}.

But as was soon pointed out in \cite{flaw}, there is a flaw in the firewall argument: the extra assumption added to make the argument work is in conflict with causality. To see this consider a black hole of mass $M$, and let the infalling object be a shell of mass $\Delta M$. When this shell reaches
\be
r'=2G(M+\Delta M)
\ee
it passes through its own horizon, and gets causally trapped inside the region $r=r'$. This part of the evolution is given by usual semiclassical physics, since we have assumed no large novel quantum gravity effects at $r>r_b$. But now the information of the shell is trapped behind a causal horizon. If we respect causality, then this information cannot escape to infinity, in violation of the assumption made in the firewall argument that Hawking radiation does carry the information of the infalling matter. One may try to argue that small quantum gravity effects may violate causality, but we already know from the small corrections theorem (\ref{one})  that small effects cannot help; we will need an order unity violation of causality. 

One may try to argue that causality does not hold in the black hole spacetime, but in that case there was no information puzzle in the first place: we can simply postulate an effect that takes the information from $r=0$ and places it, say, at $r=4GM$, after which it can escape to infinity.\footnote{The nonlocal models of \cite{giddings} are examples of how such a nonlocality model might work.}   

As we will recall below, the fuzzball paradigm does not have a similar conflict with causality because the infalling shell is expected to tunnel into fuzzballs at $r\approx r'+\epsilon$, so it never gets trapped inside its own horizon \cite{causality1, causality2}. But this implies that nontrivial quantum gravity effects happen when the new horizon is about to form in the region $r>r_b$, in conflict with the assumption of the firewall argument that the region $r>r_b$ has no nontrivial quantum gravity effects.

Given this conflict of the firewall argument with causality, one may ask a `modified firewall question': will an infalling object encounter strong interactions with emerging radiation quanta {\it before the location where it gets trapped inside its own horizon}? In the above example, this is equivalent to asking if the infalling shell of mass $M$ will get burnt before reaching $r'=2G(M+\Delta M)$,  rather than asking if it will get burnt before reaching $r=2GM+\epsilon$. 

In \cite{flaw} it was shown that the answer to the modified firewall question is negative for standard gravitational interactions. That is,   in the limit where the black hole is large
\be
{M\over m_p}\equiv C \gg 1
\label{two}
\ee
the probability that an infalling object will undergo a gravitational scattering off a radiation quantum goes to zero. Thus there is no modified firewall effect; {\it i.e., there is no firewall for large black holes if we assume that causality is not violated at leading order in the  region $r> r_b$}.

It was noted in \cite{flaw} that at high energies gravitational interactions dominate over vector and scalar interactions. So it would seem that once we have checked for the absence of modified firewall behavior for gravitational interactions, we have checked it for vector and scalar interactions as well. But it was noted by Marolf \cite{marolf} that in the standard model, there is a large dimensionless number that gives the charge to mass ratio of the electron. Thus while it is true that there will be no modified firewall behavior for sufficiently large $C$ in (\ref{two}), we should check carefully the situation for astrophysical holes: it may be that these holes are small enough that an infalling electron scatters off a radiated photon before the electron gets trapped inside its own horizon. 

We  perform this check explicitly. For the interaction of infalling electrons with outgoing photons, we find that there is no modified firewall behavior: i..e, there is a negligible chance that an electron falling into an astrophysical hole will scatter off a radiated photon before getting trapped inside its own horizon.   We will then argue that other standard model effects will not change this situation, so there is no modified firewall behavior for an infalling electron. 

\subsection{Long wavelength scattering}

The firewall argument envisages the infall of objects into the hole, and asks if these objects will get burnt by Hawking radiation before reaching the horizon. Thus if we focus on single quanta, then the wavelength of these quanta should be smaller than the radius of the black hole\footnote{We will use the symbol $r_h$ to describe the radius $r=2GM$ even in the case where we have a fuzzball in place of the traditional hole; note that with the fuzzball there is no actual horizon at this radius. Similarly, we will often use the term `black hole' even when we have in mind the fuzzball picture of black hole microstates. If we do wish to talk about a horizon, then we will call the object a `traditional hole'. }
\be
\lambda \ll r_h
\ee
We have  noted above that such quanta will turn out to  not have any significant interaction with Hawking radiation before they are swallowed by the surface of the fuzzball; i.e., there is no `modified firewall effect'. But suppose  we ask about the scattering of very {\it low} energy quanta
\be
\lambda \gtrsim r_h
\ee
We will see that photons with such long wavelengths can indeed scatter off electrons that are present in the hear horizon Hawking radiation emerging from a hot surface. But this scattering happens very close to $r=r_h$. The backscattered photons then have to be very close to the outward normal direction if they are to escape to infinity; all other backscattered photons fall back onto the surface of the fuzzball. Thus in practice it will be vary hard to detect the surface of a fuzzball.

\subsection{The plan of the paper}

The plan of the paper is as follows:

\b

(i) In section \ref{secnature} we recall the fuzzball construction. While only a subset of all possible fuzzballs  have  been constructed, we motivate some properties that the generic fuzzballs should have, in particular that their surface should be only slightly more than the radius $r=2GM$ of the Schwarzschild hole.

\b

(ii) In section \ref{secgeodesic} we show that if a geodesic starts out just outside $r=2GM$, then it cannot escape to the region far from the hole unless it is directed very close to the outward normal. This fact will limit the possibilities for observing structure very close to $r=2GM$. 

\b

(iii) In section \ref{secfirewall} we recall the argument of \cite{amps} that anything falling onto an astrophysical body with a surface like that of a fuzzball  will get burnt by the emerging radiation before reaching the surface of the fuzzball.  We then note the flaw in the firewall argument: the extra assumption required to prove firewall behavior violates causality. We then define  `modified firewall behavior' to describe a firewall that occur at a location consistent with causality.

\b

(iv) In section \ref{secinteract} we recall the computations of \cite{flaw} where it was found that for black holes of sufficiently large mass, modified firewall behavior will {\it not} exist.

\b

(v) In section \ref{secepinteract} we examine the possibility of modified firewall behavior arising for astrophysical holes when infalling electrons interact with photons in the emerging radiation. We first estimate the effect, and then compute it in detail. The interaction probability is found to be negligible, so we conclude that there is no modified firewall behavior from such interactions for astrophysical holes. 

\b

(vi) In section \ref{secgeneralanalysis} we show that using other interactions expected in standard model physics and beyond the standard model physics does not change the above conclusion; i.e., we still do not get modified firewall behavior for astrophysical holes. 

\b

(vii) In section \ref{seclow} we consider  photons  with wavelength $\lambda \gtrsim GM$ incident on the hole. These photons can scatter off electrons present in the near horizon radiation. The probability for this scattering is larger than the scattering probabilities we have found in the computations of the above sections. But the probability for the scattered photon to re-emerge to infinity is  still much less than unity for astrophysical holes, so it is difficult to detect the fuzzball surface this way as well.

\b

(viii) Section \ref{secdiscussion} is a summary and general discussion.

\section{The nature of fuzzballs}\label{secnature}

In this section we will give a heuristic picture of fuzzballs; this picture should help in understanding the scales we will use to compute scattering.  in later sections we will ask about the possibility of detecting fuzzball structure  assuming that this picture is correct.

\subsection{States in string theory}

As mentioned in the introduction, the most natural resolution to the information puzzle would be to have the black hole replaced by a ball with no horizon. The traditional problem with having such a ball is that objects that are compressed to a radius close to the horizon radius tend to undergo runaway collapse to a point, leaving behind the traditional black hole. The Buchdahl theorem \cite{buchdahl}, in particular, considers a static, spherically symmetric  ball of perfect fluid with mass $M$ and radius $R$, where the pressure $p$ assumed to increase monotonically towards the center. The gradient $dp/dr$ must support the gravitational attraction on each fluid element, so assuming that $p=0$ at the surface, we can compute $p(r)$. It is then found that if 
\be
R< {9\over 4} GM
\ee
then $p$ reaches infinity at some radius $r>0$. We  conclude that such a ball cannot support itself, and must undergo collapse to a black hole. 

String theory microstates found with the fuzzball construction are not exactly spherically symmetric;  thus, strictly speaking, the Buchdahl theorem does not apply. But we can still ask how we can avoid the {\it spirit} of the theorem, since it would seem that for $R$ sufficiently close to $2GM$ it would become very hard for the ball to resist collapse. The key point is that  string theory has novel features like extended objects and extra dimensions. In \cite{toyfuzz} a toy example was given to explain how such extra dimensions can be used nonperturbatively to get interesting solutions that do not collapse. Let us summarize this toy model. 

Consider the 4-d Euclidean Schwarzschild metric, and add in a time direction to get a 4+1 dimensional spacetime
\be
ds^2=-dt^2 + (1-{r_h\over r})d\tau^2 + {dr^2\over 1-{r_h\over r}} + r^2 (d\theta^2+\sin^2\theta d\phi^2)
\label{metrickk}
\ee
Here the `Euclidean time' direction $\tau$ is compact, with $0\le \tau < 4\pi r_h$. The spacetime ends at $r=r_h$. The overall spacetime is smooth, as the $r, \tau$ directions form a cigar, whose tip lies at $r=r_h$. 

Now consider this metric as a 3+1 dimensional solution, with the size of the direction $\tau$ giving a massless scalar
\be
g_{\tau\tau}=e^{{2\over \sqrt{3}}\Phi}
\ee
Thus the scalar field has the value 
\be
\Phi={\sqrt{3}\over 2}\ln  (1-{r_h\over r})
\label{qone}
\ee
The 3+1 dimensional Einstein metric is 
\be
g^E_{\mu\nu}=e^{{1\over \sqrt{3}}\Phi} g_{\mu\nu}
\ee
which yields
\be
ds^2_E=-(1-{r_h\over r})^\h dt^2 + {dr^2\over (1-{r_h\over r})^\h} + r^2 (1-{r_h\over r})^\h(d\theta^2+\sin^2\theta d\phi^2)
\label{metric3}
\ee
The stress tensor of the scalar field has the usual form for a massless scalar
 \be
 T_{\mu\nu}=\Phi_{,\mu}\Phi_{,\nu} -\h g^E_{\mu\nu} \Phi_{,\lambda} \Phi^{,\lambda}
 \ee
so it has a positive energy density. We now ask: what holds up this 3+1 dimensional solution from collapsing under the self-gravity of the scalar $\Phi$? 

The solution (\ref{qone}) gives the stress tensor
\be
T^\mu{}_{\nu} =  {\rm diag }\{-\rho, p_r, p_\theta, p_\phi    \}={\rm diag }\{-f, f, -f, -f    \} 
\ee
where
\be
f= {3r_h^2\over 8r^4 (1-{r_h\over r})^{3\over 2}}
\ee
We see that $\rho>0$ everywhere, and that $\rho\r \infty$ as $r\r r_h^+$. The radial pressure $p_r$ has the same behavior, diverging at $r\r r_h^+$. In the spirit of Buchdahl's theorem, we would consider such a divergence unphysical, and discard this solution. But here we see that the full 4+1 dimensional solution (\ref{metrickk}) is completely regular; it is only its decomposition into a 3+1 dimensional metric and a scalar $\Phi$ which breaks down at $r=r_h$. The region $r<r_h$ does not exist: the $\tau$ circle shrinks to zero to generate a smooth cigar in the $r, \tau$ directions with tip at $r=r_h$. Thus the overall solution has a different topology from that of 3+1 dimensional Minkowski space times a circle. 

To summarize, we can obtain new solutions in the 4+1 dimensional theory that would have been considered pathological from the 3+1 dimensional viewpoint. In string theory we have to obtain all the microstates of the black hole by looking at different bound states of branes. In each case that has been examined so far, one finds a situation similar to the toy model above where the bound state can have the allowed string theory sources (new topologies, strings, branes fluxes,  etc.) but no horizon or singularity. All 2-charge extremal states can be shown to be fuzzballs \cite{lm4,fuzzballs}. Many extremal and near extremal states  have been found to have the form described by a set  of KK monopoles and antimonopoles, with fluxes on the noncontractible spheres linking the centers of these monopoles and antimonopoles \cite{lmm,bena}.  Many classes of 3-charge extremal states have also been shown to be fuzzballs. Some families of nonextremal states have been constructed. In \cite{neutral} neutral states were constructed that had maximal rotation. All these states were found to be of fuzzball type; i.e., with no horizon or singularity. 

The fuzzballs radiate from their surface like normal warm bodies so there is no problem of growing entanglement. In some simple cases  an explicit computation of this radiation has been done \cite{radiation1}. In each of these cases one finds that  the  {\it rate} of radiation turns out to agree with that expected for Hawking radiation from the corresponding  microstate \cite{radiation2}. But this radiation does not arise from pair creation since there is no region `interior to the horizon' where negative energy particles can exist; thus the detailed correlations among the emitted particles is very different from that found in the Hawking computation \cite{hawking}, and so there is no information problem with fuzzballs.

\subsection{The size of fuzzballs}

Consider a  fuzzball of mass $M$. We assume for simplicity that the fuzzball has no charge or rotation. We will assume that the fuzzball is roughly spherical in shape. Let the outer boundary of the fuzzball be at a radius $r=r_b$. We  cannot have $r_b<2GM$, since in that case the fuzzball would be inside the horizon at $r_h=2GM$. Any object that is inside a horizon must, in  a theory that obeys causality, continue to fall in towards $r=0$. Thus we conclude that we need $r_b>2GM$.

But now we have two possibilities, which we name as follows:

\b

(i) {\it Tight fuzzballs:} We have
\be
r_b=2GM+\epsilon, ~~~\epsilon\ll GM
\ee

\b

(ii) {\it Diffuse fuzzballs:} In this case we have
\be
r_b=2GM+\epsilon, ~~~\epsilon\gtrsim GM
\label{diffuse fuzzball}
\ee

\b

We do not have a construction of generic neutral fuzzballs. Thus we cannot say for sure which of these two cases is realized. But we will now give a heuristic argument for why the case (i) is plausible; in fact we will argue that in a simple model of fuzzballs, we expect
\be
r_b=2GM+\epsilon, ~~~\epsilon\sim {l_p^2\over GM}
\ee
where $l_p$ is the planck length. This value of $\epsilon$ corresponds to $r_b$ being at a proper distance $\sim l_p$ outside the location $r_h=2GM$. The difference between (i) and (ii) is of course relevant for the observability of fuzzballs: we will see below that it is very hard to observe structure that is very close to the horizon. 

For the sake of generality, consider the Schwarzschild hole in $D$ spacetime dimensions
\be
ds^2= - \left ( 1-\left ( {r_h\over r}\right )^{D-3} \right ) dt^2 + {dr^2 \over \left ( 1-\left ( {r_h\over r}\right )^{D-3} \right )}+r^2 d\Omega_{D-2}^2
\ee
We write
\be
G\equiv l_p^{D-2}, ~~~m_p\equiv{1\over l_p}
\ee
We have
\be
GM\sim  r_h^{D-3}
\label{qfive}
\ee
and the Bekenstein entropy of the hole is
\be
S_{bek}\sim {A\over G}\sim \left ( {r_h\over l_p}\right )^{D-2}\sim \left ( {M\over m_p}\right ) ^{D-2\over D-3}
\label{qtwo}
\ee
In the near horizon region, the proper distance from the horizon is 
\be
s=\int_{r_h}^r {dr' \over \left ( 1-\left ( {r_h\over r'}\right )^{D-3} \right )^\h} \approx {2\over \sqrt{D-3}} r_h^\h (r-r_h)^\h
\label{qproper}
\ee
For $D=4$ this is
\be
s=2({2GM})^\h (r-2GM)^\h
\ee

Now suppose the fuzzball was made from elementary topological objects like KK monopoles and antimonopoles. We assume that all compact directions have size $\sim l_p$. The string coupling $g$ is also a fixed number of order unity; i.e., it does not scale with the mass of the hole. Then each elementary object making the fuzzball has a mass
\be
m\sim m_p
\ee
If we regard these elementary objects as separate entities, without any strong influence on each other, then we expect that the number of such objects in the fuzzball would be
\be
N\sim {M\over m_p}
\label{aone}
\ee
Suppose each such object carries one bit of entropy; for example the choice of being a monopole or antimonopole, or a spin contributed by fermion zero modes. Then the entropy of such fuzzballs would be
\be
S\sim N \sim {M\over m_p}
\label{qthree}
\ee
Then from (\ref{qtwo}) we see that $S\ll S_{bek}$ for $M\gg m_p$, so we do not get the full entropy of the black hole.

 A way out of this difficulty is to note that each elementary object in the fuzzball feels the gravitational potential created by all the other objects. If we have a diffuse fuzzball, then the redshift factor is $(-g_{tt})^\h \sim 1$ at generic points in the fuzzball, and we will still have an entropy  of order
(\ref{qthree}). But if we have a tight fuzzball, then at the surface $r_b$ we have a redshift
\be
(-g_{tt})^\h \sim \left ( {\epsilon\over r_h}\right )^{1/2} \sim {s\over r_h}
\label{qfour}
\ee
where in the second step we have written the redshift in terms of the proper distance from the horizon (\ref{qproper}).

If we place the elementary objects making the fuzzball at a location with this redshift, then the effective energy contributed for each such object is of order
\be
(-g_{tt})^\h m_p \ll m_p
\ee

We can now estimate the distance $s$ from the horizon where the elementary objects must be placed to get the full entropy $S_{bek}$. The number of objects we can have at redshift (\ref{qfour}) is
\be
N'\sim (-g_{tt})^{-\h}\left ( {M\over m_p}\right )
\ee
Setting the entropy $S\sim N'$ to the value $S_{bek}$, we get
\be
\left ( {r_h\over s}\right ) \left ( {M\over m_p}\right ) \sim \left ( {M\over m_p}\right ) ^{D-2\over D-3}
\ee
which using (\ref{qfive}) gives
\be
s\sim l_p
\ee
Thus if we place all the fuzzball constituents at a planck distance outside the radius $r_h$, then we can have enough constituents to get the Bekenstein entropy. For this value of $s$, we have from  (\ref{qproper})
\be
\epsilon= r_b-r_h \sim {l_p^2\over r_h}
\ee

\subsection{Summary}

In general, all we can say about the fuzzball is that its boundary $r_b$  cannot be confined inside the horizon radius $r_h$. We have however given arguments that $r_b$ should be close to $r_h$; in fact this boundary should be at a distance $\sim l_p$ outside $r_h$. We have termed such fuzzballs as `tight fuzzballs'.

Note however that the arguments in favor of tight fuzzballs are based on a picture where each of the fuzzball constituents contribute  $\sim 1$ bits  to the entropy. We can easily imagine  more complicated ways of encoding entropy, and in that case we can get `diffuse fuzzballs' whose size is given by (\ref{diffuse fuzzball}). For example, we can have `links' between the $N$ basic constituents counted in (\ref{aone}). There can be $N^2$ such links, and if each link contributed $\sim 1$ bit to the entropy, then we would get the required entropy $S\sim N^2$ for $D=4$, without needing $(-g_{tt})$ to be small anywhere.

While diffuse fuzzballs are therefore possible, we will work below with the assumption of a tight fuzzball. As remarked above, it is difficult to see structure that is very close to $r_h$. Thus if observations do indicate such structure, then such observations may point to the possibility of diffuse fuzzballs.

\section{Geodesics in the Schwarzschild metric}\label{secgeodesic}

Consider a particle that is falling into the black hole. How does its trajectory behave when it reaches very close to the horizon? If this particle scatters off something close to the horizon, what is the trajectory of the scattered particle?  Since these questions will be relevant for our computations, we begin by recalling the behavior of particle trajectories in the exterior of the Schwarzschild hole. We will now focus our attention on the 3+1 dimensional metric
\be
ds^2=-f(r) dt^2 + f(r)^{-1} dr^2 + r^2 (d\theta^2+\sin^2\theta d\phi^2)
\ee
where
\be
f(r)=1-{2GM\over r}
\ee
Consider a particle of mass $m$ moving in this metric. We can take its trajectory to lie in the equatorial plane $\theta={\pi/ 2}$. The 4-velocity $U^\mu={dx^\mu/ d\tau}$ has the nonvanishing components $U^t, U^r, U^\phi$. 
The condition $U^\mu U_\mu=-1$ gives
\be
(U^t)^2g_{tt}+(U^r)^2 g_{rr} + (U^\phi)^2g_{\phi\phi}=-1
\label{wone}
\ee
The conserved quantities are 
\be
U_t\equiv -{ E\over m}\equiv -\t E, ~~~ U_\phi\equiv { L\over m}\equiv \t L
\ee
In particular, if the particle reaches infinity, then its energy at infinity will be  $E$. 
The relation  (\ref{wone}) gives
\be
-\tilde{E}^2 f^{-1} + (U^r)^2 f^{-1} + \tilde{L}^2 r^{-2}=-1
\label{timelike geodesic}
\ee
We get
\be
U^\phi={d\phi\over d\tau}={\t L\over r^2}
\label{btwo}
\ee
\be
U^r={dr\over d\tau} = \pm \sqrt{\t E^2-{f\over r^2} \t L^2-f}\equiv \pm \sqrt{V}
\label{eone}
\ee
\be
{dr\over d\phi}= {U^r\over U^\phi}=\pm {r\over \t L} \sqrt{r^2(\t E^2-f)-{f} \t L^2}
\ee

\subsection{Escaping to infinity}\label{secescape}

Consider the expression (\ref{eone}) for $U^r$, where
\be
V=\t E^2-{f\over r^2} \t L^2-f
\ee
A particle that is outward directed will have $U^r>0$, and will continue to move outwards as long as $U^r$ remains positive. Thus if $V>0$ for all $r$ larger than the starting value $r=\bar r$, then the particle will escape to infinity. To see if $V$ becomes negative anywhere along the outward motion of the particle, we find the minimum of $V$, and check if $V$ is negative there. We find that the minimum of $V$ is at 
\be
r_{min}={\t L^2-\sqrt{\t L^4 - 12 \t L^2 (GM)^2}\over 2GM}
\ee
and the value of $V$ at this position is
\be
V_{min}=\t E^2 +{1\over 54} \left ( -36+{12\sqrt{\t L^2-12 (GM)^2}\over \t L} - {\t L (\t L + \sqrt{ \t L^2 -12 (GM)^2})\over (GM)^2}\right )
\ee
If a trajectory is at the borderline of being able to escape, then we must have $V_{min}=0$, which implies that
\be
\t E = {1\over 3\sqrt{6}} \left ( 36-{12\sqrt{\t L^2-12 (GM)^2}\over \t L} + {\t L (\t L + \sqrt{ \t L^2 -12 (GM)^2})\over (GM)^2} \right ) ^\h 
\label{rone}
\ee
In the limit where $\t L, \t E$ go to infinity (i.e., in the limit where the particle can be treated as massless) the quantity (\ref{rone}) becomes
\be
\t E = {1\over 3\sqrt{3}}{\t L\over GM}
\ee
Thus a high energy particle near the horizon will escape to infinity if its $\t E$ is larger than that above minimum value; this minimum value is determined by the $\t L$ carried by the particle.

\subsection{The escape angle in the local frame}

We now wish to convert  the above condition for escape into a condition on the initial angle $\Psi$ which the outgoing particle makes with the outward normal direction.

Consider a point at $r=\bar r$, $\theta=\pi/2$. At this point we can set up a local orthonormal frame, where
\be
d\hat t = (-g_{tt})^\h dt, ~~~ d\hat r = g_{rr}^\h dr, ~~~ d\hat \theta = \bar r d\theta, ~~~d\hat\phi = \bar r d\phi
\label{aframe}
\ee
The energy of the particle in this local frame is
\be
\hat E = f(\bar r)^{-\h} E
\ee
The radial momentum is
\be
\hat p^r = f(\bar r)^{-\h} p^r
\ee
The momentum in the direction $\hat \phi$ is
\be
\hat p^\phi = {L\over \bar r}
\ee
Suppose the particle is headed outwards at an angle $\Psi$ to the outward radial direction. Then
\be
\hat p^\phi = \hat p \sin\Psi = \sqrt{\hat E^2 - m^2} \, \sin\Psi
\ee
Thus
\be
\sin\Psi = {L\over \bar r} {1\over \sqrt{\hat E^2-m^2}}= {\t L\over \bar r} {1\over \sqrt{f^{-1}(\bar r)\t E^2  -1}}
\label{rtwo}
\ee
Our goal is to find the range of $\Psi$ where the particle is able to escape to infinity. We will see that $\Psi$ is small if the starting location $\bar r$ is close to $2GM$. 

To find $\Psi$ we do the following: (1) We solve  (\ref{rone}) to get $\t L$ in terms of $\t E$; (2) We substitute this $\t L$  in the (\ref{rtwo}) giving $\sin\Psi$. This yields the maximal value of $\Psi$ that will allow escape from the position $\bar r$ when the energy per unit mass is $\t E$. 

The result of this computation becomes simple in the limit where $\t L, \t E$ go to infinity (i.e., in the limit where the particle can be treated as massless). This will be the case of relevance to us, since we will have high energy scattering  near the horizon. In this high energy limit the quantity (\ref{rtwo}) becomes
\be
\sin\Psi = {3\sqrt{3}GM\sqrt{\bar r-2GM}\over \bar r^{3\over 2}}\approx {3\sqrt{3} \over 8} {\bar s\over GM}
\label{finalc}
\ee
where in the second step we have assumed that $\bar r-2GM\ll GM$ and written the result in terms of the proper distance $\bar s$ from the horizon. We see that if a massless particle starts at a distance $\bar s\ll GM$ from the horizon, then it must be confined within a very narrow angle $\Psi$ to the radially outward direction if it is to escape to infinity. 

\subsection{Escaping to the region $r-r_h\sim r_h$}

In the above discussion we had required that the particle escape to infinity. But what if it merely escapes the near horizon region, and is picked up at a location $r-r_h\sim r_h$? This could happen for example if we require the particle to reach the accretion disc around a hole, where it could signal its presence by interacting with the matter in the accretion disc. 

But as we will now note, relaxing the condition for escape in this way does not make much difference; we still get a very narrow angle of escape $\Psi$. Thus let the particle reach a maximum radius $r_{max}$ after which (in the absence of any other interaction)  it falls back towards the hole. At the radius $r_{max}$ we have $U^r=0$, so we get  from (\ref{timelike geodesic})
\begin{equation}
\tilde{L}^{2}r_{0}^{-2}=\tilde{E}^{2}f^{-1}(r_{max})-1
\label{bone}
\end{equation}
From (\ref{rtwo}) we get
\begin{equation}
\sin\Psi={\t L\over \bar r} {1\over \sqrt{f^{-1}(\bar r)\t E^2  -1}}=\frac{r_{0}}{\bar{r}}\frac{\sqrt{f^{-1}(r_{max})\tilde{E^{2}}-1}}{\sqrt{f^{-1}(\bar{r})\tilde{E}^{2}-1}}
\end{equation}
Note that $r_{max}>\bar r$, which gives $f^{-1}(r_{max})<f^{-1}(\bar{r})$. We then have the inequality
\be
\frac{\sqrt{f^{-1}(r_{max})\tilde{E^{2}}-1}}{\sqrt{f^{-1}(\bar{r})\tilde{E}^{2}-1}}<\frac{f^{-1/2}(r_{max})}{f^{-1/2}(\bar{r})}
\ee
We are interested in the region  $r_{max}-r_h\sim r_h$. This gives  $r_{max}\sim r_h$, and $f(r_{max})\sim 1$. We then find
\be
\sin\Psi=\frac{r_{0}}{\bar{r}}\frac{\sqrt{f^{-1}(r_{max})\tilde{E^{2}}-1}}{\sqrt{f^{-1}(\bar{r})\tilde{E}^{2}-1}}\lesssim {\bar s\over GM}
\label{finalc2}
\ee
which is the same condition as (\ref{finalc}).\footnote{Note that the relation (\ref{bone})  only a necessary condition for escape to the radius $r_{max}$; to get a necessary and sufficient condition we have to perform the minimization of the potential as done in section \ref{secescape}.}

\subsection{The angle of an infalling particle}

In the above discussion we have looked at particles that are trying to escape from a point $r=\bar r$ close to the horizon. We now consider the opposite situation. Suppose a particle is falling {\it in} from afar; i.e., from the region $r-r_h\gtrsim r_h$. Then we will see that the trajectory of this particle becomes very close to the inward normal direction when the particle reaches  a point $\bar r$ very close to the horizon. 

At any point along the infalling trajectory we have
\be
{d\phi\over dr}= {{d\phi\over d\tau}\over {dr\over d\tau}}
\ee
Let the particle be described by parameters $\t E, \t L$ as before. From (\ref{btwo}) and (\ref{eone}) we get
\be
{d\phi\over dr}=-{\t L\over r \sqrt{r^2(\t E^2-f) - \t L^2 f}}
\ee
In the local orthonormal frame (\ref{aframe}) we have
\be
{d\hat\phi\over d\hat r} = {d\phi\over dr} {g_{\phi\phi}^\h\over g_{rr}^\h}= {d\phi\over dr} r f^\h
\ee
We then get
\be
{d\hat\phi\over d\hat r} = -{\t L f^\h\over \sqrt{r^2(\t E^2-f) - \t L^2 f}}
\ee
Thus if the angle to the inward radial direction is called $\Phi$, then we have
\be
\tan\Phi = { f^\h \t L\over \sqrt{r^2(\t E^2-f)-{f} \t L^2}}
\ee
In the limit where the particle reaches a location $\bar r$ very close to the horizon, we get
\be
\tan\Phi \approx {\t L\over (2GM)^{3\over 2} \t E} (\bar r-2GM)^\h\approx {\t L\over \t E} {\bar s\over 2(2GM)^2}
\label{nearradial}
\ee
Thus we see that the particle trajectory becomes close to radial infall when $\bar s\r 0$, regardless of how the particle started. This will be relevant when we consider the infall of particles to the near horizon region; we will let the trajectory be radially inwards in the region near the horizon. 

\subsection{Summary}

Let us collect together some of the observations above to see why it will not be easy to observe the fuzzball surface:

\b

(a) One may try to observe a fuzzball surface by sending a light ray that grazes the surface of the fuzzball, with the hope that such a ray would emerge back to infinity and tell us something about the surface. But in the Schwarzschild metric we do not have particle trajectories that come in from afar, graze the region close to the horizon and emerge back to infinity. In fact, as we have seen above,  particles falling in from afar hit the fuzzball surface is a near-normal trajectory. 

\b

(b) Once a particle hits the surface of the fuzzball, we have seen from an entropy argument based on eq. (\ref{bthree}) then it is very difficult for the particle to backscatter off the surface; there is much more entropy for the particle to convert its energy into excitations of the fuzzball. 

\b

(c) If we consider the small probability that the particle does scatter off the fuzzball surface, then we face the difficulty that the scattered particle will find it very difficult to escape from the near horizon region of the hole. In fact from (\ref{finalc}),(\ref{finalc2}), we see that the solid angle around the outward normal which allows escape is
\be
\delta\Omega \sim\Psi^2 \sim \left ( {\bar s \over r_h}\right )^2
\label{escape1}
\ee
If we set $\bar s \sim l_p$ to correspond to the boundary $r_b$ of the tight fuzzball, and $r_h\sim 3\, Km$ corresponding to a solar mass hole, then we get
\be
\Delta \Omega \sim 10^{-77}
\ee
Thus if we assume that scattering off the fuzzball surface is isotropic, then there is only a probability $10^{-77}$ that the scattered quantum will escape from the near horizon region of the fuzzball. 

\b

(d) One might wonder if it is possible to see evidence of fuzzball structure in the gravitational waves emitted during black hole mergers. But we see that to get any evidence of a fuzzball surface at $r_b=r_h+\epsilon$ will be difficult, for the  same reasons as mentioned above. The computation of gravitational radiation is done {\it outside} the horizon; we simply put a boundary condition that any wave reaching the horizon is absorbed. But recall that because of the high density of states implied by (\ref{bthree}),  the fuzzball surface is highly absorptive. So this surface mimics a black hole horizon in the sense that very little reflects off this surface. Since the boundary condition at a fuzzball surface is almost the same as that at a traditional horizon, the computation of radiation is also almost the same.

Even if a small part of the wave were to reflect off the fuzzball surface, we see from (\ref{escape1}) that a negligible part of this reflected wave will emerge to the region $r-2GM\sim GM$. 

In the computation of gravitational waves using traditional holes, the part of the wave that falls inside the radius $r=3GM$ is typically absorbed by the hole and does not emerge in gravitational radiation. For the reasons given above, it appears likely that the fuzzball will mimic a similar behavior, so that it will likely be hard to see evidence of a fuzzball surface using gravitational waves. 

In \cite{cardoso1} a detailed analysis was performed of ringdown modes for black holes and for compact objects without a horizon, and it was noted that while the quasi-normal modes can be very different in the two cases, the ringdown signal is very similar.

\b

Since it is hard to observe quanta scattered off the fuzzball surface itself, one might wonder if it is possible to scatter quanta off the Hawking radiation being emitted from the hole. For incoming quanta with wavelengths $\lambda \ll  r_h$, this is the question posed in the firewall argument; we will find that there is no significant backscattering for such quanta outside the fuzzball surface. We will then consider the low energy regime $\lambda \gtrsim r_h$ which corresponds to scattering very low energy quanta ($E\lesssim T$) off the fuzzball. We will find that such quanta can backscatter off the radiation, but the backscattered quantum again has a very small probability to escape the near horizon region of the hole.

\section{Firewall behavior}\label{secfirewall}

In subsection \ref{secfire1} we  outline the AMPS argument  \cite{amps} that replacing the horizon with a radiating surface leads to a firewall of radiation just outside the horizon. In subsection \ref{secfire2} we will observe that the assumptions used in the argument are in conflict with causality. In subsection \ref{secfire3} we will define  a `modified firewall'  that would be consistent with causality. 

In later sections we will recall the argument of \cite{flaw} that such a  modified firewall does not actually arise for holes with mass $M\gg m_p$,  for a theory with standard gravitational interactions.  After that, we will also  check for modified firewall behavior involving electron-photon interactions in  realistic astrophysical holes, and again find the absence of a modified firewall. 

\subsection{Outline of the firewall argument}\label{secfire1}

Hawking's computation \cite{hawking} showed that if we have the vacuum state around the horizon, then we will get a monotonically rising entanglement between the emitted radiation and the remaining hole. Hawking's computation was a leading order computation using the semiclassical approximation around the horizon. In \cite{cern} it was proved that Hawking's argument was stable against any source of small corrections  to the evaporation process. This result is known as the `small corrections theorem', and it converts Hawking's argument of 1975 into a `theorem' that horizons will necessarily lead to information loss.\footnote{One can bypass the theorem by requiring that degrees of freedom at infinity are not independent of degrees of freedom near the hole \cite{cool, pr}; we will not be considering such possibilities here.}  We can state the theorem in an exactly equivalent way as follows. Suppose we assume 

\m

{\bf Ass:1} The information of the hole is radiated away the same way as by any other black body; i.e., there is no monotonic rise in entanglement between the radiation and the remaining hole.

\m

Then the  theorem says that {\it the horizon region cannot behave as a vacuum to leading order for low energy physics}. In other words, the picture of the horizon as a `normal place' cannot be correct; we need order unity corrections to `normal physics' at the horizon. 

\m

AMPS \cite{amps} sought to make this result stronger by adding one extra assumption, which we may state as follows:

\m

{\bf Ass:2} Consider a surface (called the `stretched horizon') placed at a location $r=r_s$ which is a planck length outside the horizon $r_h=2GM$.  Let the region $r>r_s$ (i.e., the region outside the stretched horizon) be described by `effective field theory';  i.e., we assume that  physics outside the hole is `normal physics'. In particular, if a shell is approaching the stretched horizon at the speed of light, then, by causality, the stretched horizon cannot respond to this shell  in any way until the shell actually reaches the stretched horizon.

\m

Under these assumptions, AMPS \cite{amps} argued that an infalling object will encounter radiation quanta of increasingly high energy $E_{rad}$ as it approaches the horizon, with  $E_{rad}$ reaching planck scale at  the stretched horizon. Thus not only is the region near the horizon not a vacuum, it will behave like a `firewall' for any object that is falling into the hole. 

The intuition behind the firewall argument is the following. In Hawking's computation of pair creation from the vacuum, the particles do not actually materialize until they are well separated from the horizon;   the region around the horizon remains a vacuum. Thus any actual particles (i.e. particles that can be interacted with)  are always long wavelength ($\lambda \sim r_h$) quanta. But if the radiation was emerging from a hot surface located at the stretched horizon, then one could follow these quanta back to a location close to the stretched horizon, where they will have a high energy due to blueshifting in the gravitational field of the hole. They will still be real particles at this location, and so they can interact with and burn an infalling object. 

This rough intuition can be made precise using the kind of bit model for Hawking radiation that was used in the small corrections theorem \cite{cern}, and a crisp argument along these lines was made in \cite{amps}.

\subsection{The difficulty with the firewall argument}  \label{secfire2}

We  see however that there is a problem with the firewall argument as it stands, since Assumptions Ass:1 and Ass:2 are in conflict with each other in any gravity theory that obeys causality \cite{flaw}. The problem can be seen  as follows:

\m

 (a) Consider a black hole of mass $M$. By assumption Ass:2,   the region $r>r_s$ is described by  `normal physics', given by usual effective field theory.

\m

 (b) Now consider a shell of radially ingoing gravitons, carrying a total energy $\Delta M$.
 Since this shell moves at the speed of light, it continues to move inwards all the way to $r=r_s$, with a dynamics governed just  by effective field theory (again by assumption Ass:2).
 
\m
 
  (c) The total mass of the black hole and the shell is $M+\Delta M$, which corresponds to a horizon at the location,
\be
r'=2G(M+\Delta M)
\label{qttone}
\ee
From (a) we see that the shell must pass `without drama' through the location (\ref{qttone}). But then  the information in the shell is trapped inside its own horizon, and cannot reach infinity unless we have a violation of causality.

\m

(d)  But we cannot violate causality in the region $r>r_s$, since this region is assumed to be described by effective field theory. Thus we find that the information in the shell {\it cannot} be  radiated to infinity, in violation of assumption Ass:1.\footnote{We can of course let the information be trapped in a remnant, and then perhaps leak out very slowly over a timescale much longer than Hawking evaporation time. But this is not what was assumed in Ass:1 -- this assumption was  really requiring the hole to radiate like a normal body and send its information out in the radiation.} 

\m

Note that if we are willing to violate causality, then there is no information puzzle in the first place; we can always say that some mechanism takes the information from the singularity and puts it outside the hole. Thus we see that causality creates a conflict between the assumptions Ass:1 and Ass:2  used in the firewall argument; in consequence we cannot argue that black hole horizons must act like firewalls. 

The goal of the AMPS argument was to argue that one cannot have {\it complementarity} \cite{complementarity}; i.e., there can be no effective dynamics of the region inside the horizon which reproduces `infall without drama' through the horizon. But once we modify the assumptions to respect causality, we can actually make a bit model \cite{bitmodel} that yields
{\it fuzzball complementarity}\cite{fuzzcomp,causality1}; i.e. infall though the horizon without drama for quanta with energies $E\gg T$.

Note also that one cannot get the information out to infinity through small quantum gravity corrections to the effective field theory assumed to hold in the region $r>r_s$. The small corrections theorem \cite{cern} says corrections of order $\epsilon$ to the state of each pair  can recover only a fraction $\sim \epsilon$ of the information; one cannot encode more information in `subtle' correlations between the emitted particles. 

\subsection{The modified firewall conjecture}\label{secfire3}

Given the above problem with causality, we may make a `modified firewall conjecture' as follows:

\b

(i) Consider a particle with energy $E$ falling towards a  black hole of mass $M$. 

\m

(ii) At a certain distance $s_{bubble}$ from the horizon, semi-classical physics says that the particle will be swallowed by a new horizon. (This new horizon will have  a bubble shaped distortion compared to the original horizon, hence the subscript `bubble'.) The distortion of the horizon is created by the backreaction of the extra energy $E$ carried by the particle. 

\m

(iii) Therefore if we are to avoid a conflict with causality, new quantum gravitational effects must arise at or before this location $s_{bubble}$ is reached. In the fuzzball paradigm, a tunneling into fuzzballs ensues just before the particle reaches this location $s_{bubble}$ \cite{tunnel}, so there is no conflict with causality. 

\m

(iv) We can still assume that semiclassical physics holds in the region $s>s_{bubble}$, and apply the firewall argument to this region. Thus we look for interactions of the infalling particle with the emerging radiation in the region
\be
s_{bubble}<s<\infty
\label{qregion}
\ee
If the probability for such an interaction is
\be
P_{interact}\sim 1
\ee 
then we have `modified firewall behavior'. If on the other hand we have
\be
P_{interact}\ll 1
\ee
then we have {\it no} modified firewall behavior. 

Let us now estimate $s_{bubble}$. The detailed deformation of the horizon upon infall of a particle can be found in \cite{virmani}. But the rough scale for $s_{bubble}$ can be found by assuming that the deformation of the horizon is in the shape of a hemisphere of radius $s_{bubble}$ \cite{flaw}. 

The basic thermodynamic relation describing the hole is
\be
TdS_{bek}=dE
\ee
Thus
\be 
\delta S_{bek} \sim {E\over T}
\ee
On the other hand we have
\be
\delta S_{bek} \sim {\delta A\over G}
\ee
where $\delta A$ is the increase in the area of the horizon. Assuming the hemispherical deformation mentioned above, we have
\be
\delta A \sim s_{bubble}^{D-2}
\ee
where the spacetime dimension is $D$. Thus we have
\be
{s_{bubble}^{D-2}\over l_p^{D-2}} \sim {E\over T}
\ee
which gives
\be
s_{bubble}\sim  \left ( {E\over T}  \right ) ^{1\over D-2} l_p
\label{ssbubble}
\ee
Our goal will be to compute the probability $P_{interact}$ for the infalling quantum to interact with a quantum of Hawking radiation in the  region $s>s_{bubble}$.

\section{Estimating $P_{interact}$ for modified firewall behavior}\label{secinteract}

In subsection \ref{secest1} we set up the scales involved in the black hole and its radiation. In subsection \ref{secest2} we recall the estimate of $P_{interact}$ preformed in \cite{flaw}. This estimate  shows the absence of modified firewall behavior for gravitational scattering, assuming the high energy cross section defined by `tiny black hole formation' as the leading interaction process.  In subsection \ref{secest3} we compute the interaction assuming a naive scattering cross section based on tree level scattering, and observe that this gives the same result in 3+1 dimensions. 

In the next section we will consider   electron-photon scattering,  and again find the absence of modified firewall behavior.

\subsection{The scales of the black hole}\label{secest1}

We will work in a general spacetime dimension $D$, and set $D=4$ at the end. Thus the metric of the hole is
\be
ds^2= - \left ( 1-\left ( {r_h\over r}\right )^{D-3} \right ) dt^2 + {dr^2 \over \left ( 1-\left ( {r_h\over r}\right )^{D-3} \right )}+r^2 d\Omega_{D-2}^2
\ee
We have $G\equiv l_p^{D-2}$. The mass and temperature are of order
\be
M\sim {r_h^{D-3}\over l_p^{D-2}}, ~~~T\sim {1\over r_h}
\ee
In the near horizon region $r-r_h\ll r_h$, the proper distance from the horizon is, from   (\ref{qproper})
\be
s\sim r_h^\h (r-r_h)^\h
\ee
At any location outside the horizon, we can set up a local orthonormal frame (\ref{aframe}). In this frame,  we will denote quantities by a hat. The local temperature and photon number density are
\be
\hat T \sim {1\over s}, ~~~\hat n \sim {1\over s^{D-1}}
\ee
The energy of a typical Hawking radiation quantum is
\be
\hat E_{radiation}\sim \hat T \sim {1\over s}
\label{qrad}
\ee
The redshift factor at this location is
\be
(-g_{tt})^\h \sim {s\over r_h}
\ee

Consider an infalling quantum which had energy $E$ at infinity. When this quantum reaches a point at distance $s$ from the horizon, its energy in the local orthonormal frame is
\be
\hat E_{infalling}\sim (g_{tt})^{-\h} E \sim E {r_h\over s} \sim \left ( {E\over T}  \right ) {1\over s}
\label{qinfall}
\ee

For the  estimates of this section,  we will assume that the Hawking radiation particles are massless. Two of the particles of interest are gravitons and photons, and these are indeed massless. We will also consider the infall of an electron,  but  when the infalling electron reaches near the horizon, it is traveling very fast 
when seen from the viewpoint of an emerging Hawking quantum, and so it can again be regarded as massless. 

In later sections  when we do a more precise computation of electron-photon scattering, we will keep the mass $m$ explicitly in the computations.

\subsection{Scattering cross section for black hole formation}\label{secest2}

Consider a graviton that falls radially into the hole, starting with energy $E$ at infinity. Assume that this graviton collides with a radially outgoing  graviton which is a quantum of Hawking radiation.

If this collision takes place close to the horizon, then it will be  a high energy collision; this is the case because the infalling graviton is blueshifted to high energies near the horizon.  It was argued in \cite{banks} that in high energy graviton-graviton collisions, the dominant interaction process results in  the formation of a tiny black hole. In this subsection we will assume that this is indeed the relevant collision process, and compute the probability $P_{interact}(s)$ for the infalling graviton to collide with a graviton of Hawking radiation by the time it has reached  a distance $s$ from the horizon. 

We begin by estimating the cross section for the collision. We first go to the center-of-mass frame, where the two interacting gravitons have the same energy $E_{cm}/2$. The collision of such gravitons will form a tiny black hole of radius
\be
r_{bh}\sim (G E_{cm})^{1\over D-3}
\label{sone}
\ee
The black hole is expected to form if the impact parameter $b$ satisfies $b\lesssim r_{bh}$. Thus the interaction cross section is
\be
\sigma_{bh}\sim (G E_{cm})^{D-2\over D-3}
\label{sigmabh}
\ee

Now let us estimate $E_{cm}$. Suppose the collision happens at a distance $s$ from the horizon. We assume that $s\ll GM$, since far from the horizon the density of Hawking quanta is very low. In a local orthonormal frame at location $s$,  the infalling graviton has energy (\ref{qinfall}) and the outgoing Hawking quantum has energy (\ref{qrad}). We assume that $E\gg T$, so that the infalling graviton has much more energy than the outgoing quantum. The center-of-mass frame is one that is  boosted in the radially ingoing direction with a boost factor 
\be
\gamma_{boost}\sim \left ( {\hat E_{infalling}\over \hat E_{radiation}}\right ) ^\h \sim \left ( {E\over T}  \right ) ^\h
\ee
With this boost, we find
\be
E_{cm}\sim \hat E_{radiation}\sim \hat E_{infalling}\sim {1\over s} \left ( {E\over T}  \right ) ^\h
\ee
Using (\ref{sigmabh})  we get
\be
\sigma_{bh}\sim \left ( G{1\over s} \left ( {E\over T}  \right ) ^\h \right ) ^{D-2\over D-3}
\ee

We now wish to find the probability $P_{interact}(s)$ that the infalling graviton interacts with a Hawking quantum up to the time that it has fallen to the location $s$. For this computation we need the number density of quanta in the Hawking radiation
\be
\hat n(s) \sim \big(\hat T(s) \big)^{D-1} \sim {1\over s^{D-1}}
\ee
We then have for the differential probability of interaction
\be
{dP_{interact}(s)\over ds} \sim  \sigma (s) \hat n(s)  \sim \left ( G{1\over s} \left ( {E\over T}  \right ) ^\h \right ) ^{D-2\over D-3}\!{1\over s^{D-1}}
\ee
Integrating this gives
\be
P_{interact}(s) \sim \left ( G \left ( {E\over T}  \right ) ^\h \right ) ^{D-2\over D-3} \!{1\over s^{(D-2)^2\over D-3}}
\ee
Setting $P_{interact}(s)\sim 1$ gives the location 
\be
s_{interact}\sim  \left ( {E\over T}  \right ) ^{1\over 2(D-2)}l_p 
\label{ss}
\ee
Thus if the infalling graviton falls in the semiclassical background of the hole upto the value of $s_{interact}$ given by (\ref{ss}), then it will have a significant probability of scattering with a Hawking quantum. But we must compare this value $s_{interact}$ with $s_{bubble}$ given in (\ref{ssbubble}), since at the location $s_{bubble}$ new quantum gravitational physics must set in to avoid horizon formation and the consequent loss of causality. We find that
\be
{s_{bubble}\over s_{interact}} \sim \left ( {E\over T}  \right ) ^{1\over 2(D-2)}
\ee
For $E\gg T$, we see that
\be
{s_{bubble}\over s_{interact}} \gg 1
\ee
Thus there is no `modified firewall behavior'. To recap, the incoming quantum gets swallowed by the deformed horizon before it can get to a region where the Hawking radiation is strong enough to give a significant probability of interaction. To solve the information puzzle without violating causality, one must invoke new quantum gravity effects at or outside the location $s_{bubble}$. By contrast the original firewall argument of \cite{amps} explicitly assumed that no novel quantum gravity effects could arise  outside the {\it undeformed} horizon. Thus the problem with the firewall argument can be traced to the fact that the backreaction of the infalling quantum on the geometry was not considered in the argument.

\subsection{Naive scattering}\label{secest3}

In the above investigation of modified firewall behavior, we assumed that the high energy interaction of gravitons was governed  by the formation of tiny black holes. It has been argued that is indeed the correct process in high energy collisions \cite{banks}. But for completeness we also check the behavior we would get if we used a more naive expression for the scattering cross section that simply extrapolates the  low energy scattering cross section. On dimensional grounds, this `naive' scattering cross section has the form
\be
\sigma_{naive}\sim G^2 (\hat E_{radiation} \hat E_{infalling})^{D-2\over 2}
\ee
Following the same procedure as above, we get
\be
{dP_{interact}^{naive}(s)\over ds}\sim \sigma_{naive}(s) \hat n(s) \sim G^2 \left ( {1\over s^2}  \left ( {E\over T}  \right ) \right ) ^{D-2\over 2}\!{1\over s^{D-1}}\sim G^2  \left ( {E\over T}  \right ) ^{D-2\over 2}\!{1\over s^{2D-3}}
\ee
This gives
\be
P_{interact}^{naive}(s) \sim  G^2  \left ( {E\over T}  \right ) ^{D-2\over 2}\!{1\over s^{2D-4}}
\ee
We see that $P_{interact}^{naive}(s)\sim 1$ for
\be
s^{naive}_{interact}\sim    \left ( {E\over T}  \right ) ^{1\over 4} l_p
\label{sbh}
\ee
We see that for $D=4$ this is the same as (\ref{ss}).  

For $D>4$, $s_{interact}^{naive}>s_{interact}$, and in fact for $D\ge 6$ we get
$s_{interact}^{naive}>s_{bubble}$. One may then wonder if there can be a modified firewall effect for $D\ge 6$. But as we now note, the naive scattering cross section $\sigma_{naive}$ actually describes for the most part just  weak gravitational deflections rather than hard scattering. To see this, note that this cross section corresponds to an impact parameter
\be
b\sim \sigma^{1\over D-2}_{naive} \sim l_p^2 (\hat E_{radiation} \hat E_{infalling})^\h
\ee
We can write this as
\be
b_{naive}\sim l_p (\hat E_{cm} l_p)
\ee
Recall that for black hole formation, the impact parameter was 
\be
b_{bh}\sim r_{bh}\sim (GM)^{1\over D-3}\sim (l_p \hat E_{cm})^{1\over D-3}\,  l_p
\ee
For $b\sim r_{bh}$ we get strong gravitational effects (including the possibility of black hole formation), but for $b\gg r_{bh}$ we get only very weak deflections: the incoming particle does not flip spin and changes its direction by a very small angle. Such deflections thus correspond to small, smooth corrections to the background metric, rather than hard scattering. It therefore does not appear that we can get a modified firewall effect even for $D\ge 6$.

\section{Electron-photon scattering}\label{secepinteract}
 
 In the last section we studied graviton-graviton scattering to check for a modified firewall effect. We found that there is no such effect: in the limit $E\gg T$ for the infalling graviton, the probability $P_{interact}$ for the infalling graviton to interact with a Hawking radiation graviton outside the deformed horizon goes to zero as a power of $T/E$. 
 
Gravitons are tensor objects of spin 2.  In principle we should also check this computation for scattering of  vector and scalar quanta. But it was noted in \cite{flaw} that at high energies the interaction cross sections for vectors and scalars grows less rapidly than the cross section for gravitons. In the limit of large black holes $M/m_p\r \infty$, we get $\hat E(s)\r\infty $, where $\hat E(s)$ is the energy of the infalling object measured in a local orthonormal frame at a distance $s$ from the horizon. Thus if there is no modified firewall effect in the limit $M/m_p\r\infty$ for gravitons, then there cannot be one for vector or scalar interactions either. 

Astrophysical black holes typically have $M/m_p\gg 1$, so one would imagine that the modified firewall effects would indeed be ignorable. But it was pointed out by Marolf \cite{marolf} that for electron-photon interactions, one should check   the condition $M/m_p\gg 1$ more carefully. If we take a toroidal compactification of string theory, with all compact directions of order the planck scale, then all we need for the above discussion to hold is that $M/m_p$ be much larger than unity. But the standard model has an unexplained hierarchy of scales: the electromagnetic interaction between electrons is  $10^{40}$ times the gravitational interaction. It may be that in this situation the condition $M/m_p\gg 1$ should be replaced by $M/m_p\gg 10^{40}$, or some other such similar condition. In that case it would still be true that there is no modified firewall behavior for {\it sufficiently} large holes, but it may be that many {\it astrophysical} holes are not large enough to reach the limit where modified firewall behavior becomes ignorable. 

We will now consider an electron that falls in from infinity, and scatters off a photon of Hawking radiation. In subsection \ref{secep1} we will perform an estimate of the probability of interaction $P_{interact}^{e\gamma}$ for this interaction along the same lines that we followed in the last section for graviton-graviton scattering. We will again find that modified firewall behavior is negligible for astrophysical holes. In subsection \ref{secep2} we will then set up the computation of scattering in full detail, since electron-photon scattering can be a useful probe of near horizon physics in a variety of scenarios.

\subsection{An estimate of electon-photon scattering}\label{secep1}

We now set the spacetime dimension to  $D=4$. On dimensional grounds, we may estimate the electron-photon scattering cross section as
\be
\sigma_{e\gamma}\sim {\alpha^2\over  E_{cm}^2}
\label{qep}
\ee 
Following the same steps as in  section \ref{secest2}, 
we have
\be
 E_{cm} \sim {1\over s} \left ( {E\over T}  \right ) ^\h
\label{COM}
\ee
Thus
\be
\sigma_{e\gamma} \sim {\alpha^2 s^2} \left ( {E\over T}\right )^{-1}
\label{qep 2}
\ee
Thus we get
\be
{dP_{interact}^{e\gamma}\over ds} \sim  \sigma_{e\gamma} \hat n\sim {\alpha^2} \left ( {E\over T}\right )^{-1} {1\over s}
\label{dpds estimate}
\ee
Integrating this gives
\be
P_{interact}^{e\gamma}(s) \sim {\alpha^2} \left ( {E\over T}\right )^{-1} \ln {r_h\over s}
\ee
Setting $P_{interact}^{e\gamma}(s)\sim 1$ gives
\be
s_{interact}^{e\gamma}\sim  r_h e^{-{1\over \alpha^2}\left( {E\over T}\right)}
\label{interaction distance}
\ee
To check for modified firewall behavior, we should compare this to
\be
s_{bubble}\sim  \left ( {E\over T}  \right ) ^{1\over 2} l_p
\label{sbubble}
\ee

For an electron that falls in from infinity, we have
\be
E\ge m
\ee
We also have $T\sim 1/r_h$. Thus we have
\be
s_{interact}^{e\gamma} < r_h e^{-{1\over \alpha^2}\left( {mr_h}\right)}
\ee
while
\be
s_{bubble} > (mr_h)^\h l_p
\ee
It is convenient to write the mass $m$ in terms of the Compton wavelength of the electron
\be
\lambda_c\sim {1\over m}
\ee
Then we get
\be
s_{interact}^{e\gamma} < r_h e^{-{1\over \alpha^2}\left( {r_h\over \lambda_c}\right)}
\label{epinteract}
\ee
\be
s_{bubble} > \left ({r_h\over \lambda_c}\right )^\h l_p
\ee
Note that 
\be
\alpha \approx {1\over 137} <1
\ee
Also, for an astrophysical hole
\be
\lambda_c \ll r_h
\ee
We see that for typical holes,
\be
s_{bubble} \gg s_{interact}^{e\gamma}
\ee
and so we again find no modified behavior if we take the interaction (\ref{qep}). As an example, consider a solar mass hole, which has $r_h\sim 3\times 10^3  \, m$. With
\be
\lambda_c\sim  \, 10^{-12} \,m,~~~ l_p\sim10^{-35} \, m
\ee
we find
\be
s_{interact}^{e\gamma}\sim 10^{-10^{16}}\, m, ~~~s_{bubble}\sim  10^{-27}\, m
\ee

\subsection{Electron Scattered by the Hawking radiation}\label{secep2}

Let us perform a full computation for infalling electrons scattered by photons in the  radiation bath.
The  scattering in the black hole background can be approximated by a scattering in the locally flat patch of space where the scattering occurs. We have computed the scattering probability in this locally flat patch (Minkowski space) in Appendix (\ref{P appendix}). 

Here we apply that result to the local orthonormal frame (\ref{aframe}) oriented along the Schwarzschild directions $(t, r, \theta, \phi)$. Let the infalling electron be moving in the negative $r$ direction with a velocity of magnitude $\beta$; we define $\gamma=1/\sqrt{1-\beta^2}$. Let the infalling electron collide with a photon of energy $\hat \omega$ in the thermal bath. 
  
Then  we find
\bea
{dP\over ds} \!\!\!&=&\!\!\!  \frac{\alpha^2}{\pi } \int   {1\over \gamma^2\beta^2}\frac{1}{e^{\hat \omega/{\hat T}}-1}  d \hat \omega\cr
&&~~\times \int_{{\gamma \hat \omega\over m}(1 - \beta)}^{{\gamma \hat \omega\over m}(1+\beta)}    \Big(\frac{1+x}{x^{2}}\Big[\frac{2x(1+x)}{1+2x}-\ln(1+2x)\Big]+\frac{1}{2}\ln(1+2x)-\frac{x(1+3x)}{(1+2x)^{2}}\Big)   dx  \nn
\label{maineq}
\eea
  
In the computation of  ${dP\over ds} $ we see that we need the two limits of the $x$ integral. To compute these limits, we put the above scattering process in the context of the black hole.
We write the black hole metric as
\be
ds^2=-f dt^2 + f^{-1} dr^2 + r^2 d\Omega^2
\ee
where $f=1-{2GM\over r}$. In the region $s\ll r_h$, we have
\bea
f(s)&\approx &\left(\frac{s}{4GM}\right)^{2}
\eea
The energy of the infalling electron in the local orthonormal frame oriented along the Schwarzschild directions is
\be
\hat{E}=E f^{-1/2} \approx {4GME\over s}
\ee
The boost parameter $\gamma$ of the infalling electron in the near horizon region is given by  
\be
\gamma(s)= {\hat E\over m} \approx \frac{4GME}{sm}
\ee
which gives
\be
\beta(s) \approx 1-\frac{1}{2}\left(\frac{sm}{4GME}\right)^{2}
\ee
The temperature in the local orthonormal frame is
\bea
\hat{T}(s)&\approx& \frac{1}{2\pi s}
\label{sch frame near}
\eea

 We can now estimate the two limits of the integral in (\ref{maineq}). For the upper limit we find 
\begin{equation}
\frac{\gamma \hat\omega (1+\beta)}{m}\sim\frac{G M}{ms^{2}}\frac{E}{m}\gg 1
\end{equation}
For the lower limit we find
\begin{equation}
\frac{\gamma \hat\omega (1-\beta)}{m} \sim\frac{1}{G mM}\frac{m}{E}\ll 1
\end{equation}
From the expressions (\ref{Flow}) and (\ref{Fhigh}) in the Appendix, we see that the lower limit gives a finite contribution, while the upper limit gives a contribution that grows  as $1/s^{2}$. Thus
in our situation where the scattering happens at small $s$, we keep only the contribution from the upper limit of the integral. Then  the relation (\ref{maineq}) becomes
\begin{equation}\label{pclose}
{dP^{e\gamma}_{interact}\over ds}=\frac{ \alpha^{2}}{\pi}\int {1\over \gamma^{2}}F_{\infty}(2\frac{\gamma \hat\omega}{m})\frac{1}{e^{\hat\omega/\hat{T}}-1}d\hat\omega
\end{equation}

\subsection{Comparing with the rough estimate}

In section \ref{secep1} we used a rough estimate of the electron-photon scattering cross section to conclude that a modified firewall did not exist for macroscopic holes. Let us check that the actual scattering cross section (\ref{pclose}) agrees with the rough estimate.

 In the expression (\ref{pclose}) we set  $\hat\omega\sim \hat{T}$. This gives $\frac{1}{e^{\hat\omega/\hat{T}}-1}\sim 1$. We also set $\int d\hat\omega\sim \hat{T}$. Using the approximation for $F_{\infty}$ in (\ref{Fhigh}), we find
 \begin{equation}
{dP^{e\gamma}_{interact}\over ds} \sim {\alpha^2 \over\gamma}{\hat{T}^2\over m}\log({\gamma \hat{T}\over m})
\end{equation}
We set the log term to unity since there is a more rapid variation from other factors in the result. Using the expressions for  $\gamma$ and $\hat{T}$  in (\ref{sch frame near}) yields
\begin{equation}
\begin{split}
{dP^{e\gamma}_{interact}\over ds}~\sim \frac{ \alpha^{2}}{ GME}\frac{1}{s}
\label{dpds 2}
\end{split}
\end{equation}
Our earlier rough estimate for 
 ${dP^{e\gamma}_{interact}\over ds}$ given in (\ref{dpds estimate})  had the form
\begin{equation}
\frac{dP^{e\gamma}_{interact}}{ds}\sim \alpha^{2}\left(\frac{E}{T}\right)^{-1}\frac{1}{s}
\sim\frac{\alpha^{2}}{GME}\frac{1}{s}
\end{equation}
which agrees with (\ref{dpds 2}).

\subsection{The computation}

  Let us now return to the full expression  (\ref{pclose}), and see what it yields for the scattering probability for a solar mass black hole. 

\begin{figure}
  \centering
    \centering
    \includegraphics[width=4in]{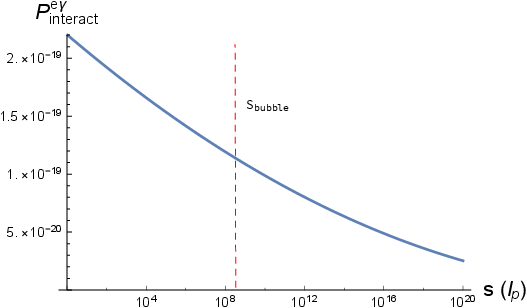}
    \caption{The probability $P^{e\gamma}_{interact}$ for an infalling electron to interact with photons in the radiation. The electron is taken to fall into a solar mass hole, starting  from rest at infinity.  The horizontal axis gives the distance $s$ from the horizon in planck units. We see that $P^{e\gamma}_{interact}$ remains small all the way from infinity to the location $s_{bubble}$. }
    \label{close}
\end{figure}

We have seen that the scattering cross section for electron-photon scattering decreases with energy. Thus to get the largest possible scattering, we let the electron start at rest at infinity; i.e.,  we set
\be
E=m
\ee
Consider a solar mass black hole.  It is convenient to write all quantities in planck units
\bea
M&=&0.928\times 10^{38}\, m_p\cr
m&=&4.186\times 10^{-23}\, m_p\cr
\alpha&=&\frac{1}{137}
\label{solarmass}
\eea
The probability of interaction $P^{e\gamma}_{interact}$ for this case is plotted in Fig.\ref{close}.  In this computation we have used $s_{max}=10^{30}l_p$; the precise value of $s_{max}$ is not important since the contribution from the region of large $s$ is negligible. The location  $s_{bubble}$ is given by 
(\ref{sbubble}). If we set $T=1/(8\pi GM)$ and 
$E=m$, then for  a solar mass black hole $s_{bubble}$ becomes
\bea
s_{bubble}\approx 3.1\times 10^8\, l_p
\label{sbubble2}
\eea
The vertical red line in fig.\ref{close} gives the location $s_{bubble}$ for the electron falling from rest. At locations $s<s_{bubble}$ the infalling electron would be already swallowed by the fuzzball surface, so we cannot look for modified firewall behavior at $s<s_{bubble}$.

In the region $s>s_{bubble}$ we see that 
\be
P^{e\gamma}_{interact}\ll 1
\ee
In fact the largest value for $P^{e\gamma}_{interact}$ takes place at $s=s_{bubble}$, and here the value is
\be
P^{e\gamma}_{s_{bubble}}\approx 1.14\times 10^{-19}\ll 1
\ee
Thus we do not get any modified firewall in this setting; i.e. for electrons falling from rest into a solar mass hole and interacting with photons in the emerging radiation. 

Figure\, \ref{close} shows the numerical results for the integration of (\ref{pclose}) in the range $s\in ( l_p, 10^{9}l_p)$. We see that the interaction probability never reaches order one outside the bubble; in fact it is very small.\footnote{Even if we go inside $s_{bubble}$ see find that at the fuzzball surface at $s=l_p$, the interaction probability is $P^{e\gamma}_{surface}\approx 2.2\times 10^{-19}\ll 1$.} Thus we do not get a modified firewall from infalling electrons interacting with photons, for solar mass holes. A similar conclusion holds for holes of mass $\sim10^8$ times solar mass which are to be found at the centers of galaxies; the interaction probabilities are even smaller for these larger holes.

\section{A general analysis}\label{secgeneralanalysis}

We have investigated the possibility of a `modified firewall effect' in two very different limits. One was the very high energy limit, where the collision of the incoming particle with a Hawking quantum results in the formation of a mini black hole. The other limit corresponded to Compton scattering of an infalling electron with a Hawking photon. In neither case did we find the modified firewall effect for macroscopic holes. But there are many energy scales between these two limits, and one might wonder whether the physics encountered at those scales could lead to a different conclusion. 

In this section we discuss what might happen at these intermediate scales. We do not of course know all the physics between the TeV scale and the planck scale. But a general analysis based on what we do know does not suggest a change in the above conclusion. We will see that the effects of many other physical processes  is similar to the effect of Compton scattering, and the relevant cross sections can be estimated in a similar way.

We start with the energy scale of simple tree level Compton scattering, and consider what new effects  arise as we increase the energy scale.

\subsection{Running of $\alpha$}

The first effect we have to note is that the electromagnetic coupling runs with energy. The effective electromagnetic coupling constant, $\alpha_{eff}$, as a function of scale $q$ for energies $-q^2\gg m^2$ is given by
\bea
\alpha_{eff} = {\alpha\over 1-{\alpha\over 3 \pi}\log({ -q^2 \over A m^2})}
\eea
where $A = e^{5/3}$.

Consider the scale where $\alpha_{eff}$ increases to unity according to this expression. For an electron of mass $m$ falling from rest at infinity, we have
\be
\sqrt{-q^2}\sim \hat E_{cm}={1\over s}\sqrt{8\pi GMm}
\ee
We then find that $\alpha_{eff}=1$ if the interaction takes place at a distance $s$ from the horizon, where $s$ is given by 
\bea
 s_{\alpha_{eff} =1}=\sqrt{{8 \pi M\over Am e^{3\pi({1 - \alpha\over \alpha})}}}~ l_p
\eea
For a solar mass hole (\ref{solarmass}), we find 
\be
 s_{\alpha_{eff} =1} \approx 10^{-248}l_p\ll l_p
\ee
Since $s_{bubble}\gg l_p$, we see that 
 there is no significant effect of the running of $\alpha$ in our computation of Compton scattering. (The Landau pole where $\alpha_{eff}=\infty$ will occur even closer to the horizon (at $s\approx 10^{-250} l_p$ for a solar mass hole).)

\subsection{Interactions with other particles}

In our analysis of Compton scattering, we have assumed that the infalling electron collides with a photon of Hawking radiation,. But as we get closer to the horizon, the local temperature rises, and particles other than photons will be present in the thermal bath. Let us look at some of the scattering processes that can result from these other particles in the bath.

\subsubsection{Electron-electron scattering}

Suppose the thermal bath contains electrons. Then we have the process $e^{-}+e^{-}\rightarrow e^{-}+e^{-}$ mediated by a virtual photon. This has a cross section at tree level
\begin{equation}\label{cee}
\sigma_{e^{-}e^{-}}\sim\frac{\alpha^{2}}{  E^{2}_{cm}}
\end{equation}
which is of the same order as Compton scattering. At high temperatures the number density of electrons will be of the same order as the number of photons, so the interaction probability from this scattering will be similar to what we have estimated for Compton scattering.

A similar analysis holds for scattering off positrons: $e^{-}+e^{+}\rightarrow  e^{-}+e^{+}$:
\begin{equation}
\sigma_{e^{-}e^{+}}\sim\frac{\alpha^{2}}{  E^{2}_{cm}}
\label{bseven}
\end{equation}
Note that the total cross-section of such processes diverges since the exchanged particle (the photon) is massless. This divergence  however arises from scattering in the near-forward direction. This  near-forward scattering  corresponds to weak deflections arising from long distance electromagnetic forces, and therefore does not describe hard scattering. In particular, if the infalling electron scatters through a small angle, then it will continue to fall in instead of emerging back to infinity. Thus the relevant part of the cross-section is indeed the finite quantity (\ref{bseven}).

\subsection{Energy Scales between tree level Compton scattering and Black Hole Formation}

We have seen in section \ref{secest2} that very high energy collisions should result in the formation of mini black holes. We have also studied the simple process of Compton scattering, which should dominate at lower energies. But in between these two energy scales are several other scales which we should note. For each of these energy scales $E_{scale}$, we can find the distance $s$ at which the center of mass energy of the collision with the infalling electron  will become $\sim E_{scale}$. 

For the electroweak scale, $E_{EW}=246$ GeV, the distance is $s_{EW}\approx 10^{25} l_p$.  For $E_{QCD} = 217$ MeV, we get $s_{QCD}\approx 10^{28} l_p$. If we take the SUSY breaking scale as  $E_{SUSY} \approx 1$ TeV, then the distance from the horizon is $s_{SUSY}\approx  10^{24} l_p$.  For the GUT scale $E_{GUT}\approx 10^{16}$ GeV  we get $s_{GUT}\approx 
10^{11} l_p$. 

Comparing these locations to $s_{bubble}$ (\ref{sbubble2}), we see that
\bea
s_{bubble} < s_{GUT} < s_{SUSY} < s_{EW} < s_{QCD}
\eea
so we need to consider interactions of the infalling electron with the particles we encounter in the thermal bath at each of these scales.

But we expect that the nature of the interaction cross section remains the same; i.e.,  the high energy interaction cross section has the form
\be
\sigma \sim \frac{1}{  E^{2}_{cm}}
\label{bsevenp}
\ee
If we assume that the number of relevant particle species in our theory is finite and not unnaturally large, then we will not find a modified firewall from such  interactions either.

\subsection{Including gravity}

We now turn to processes which include gravity. Since the coupling constant for gravity $G_N\sim l_p^2$ has units of $(length)^2$, the cross section will have positive powers of $E_{cm}$ compared to the standard model interactions above.  For this reason it is important to consider the maximum value of the energy  at which we are looking at these interactions. 

For a modified firewall, we are looking for an order unity probability for interaction outside the location $s_{bubble}$. Recall that
\begin{equation}
s_{bubble}\sim \left(\frac{E}{T}\right)^{1/2}l_p
\end{equation}
If the collision with a radiation quantum happens at the location $s$, then  the center of mass energy (\ref{COM}) is
\begin{equation}
E_{cm}\sim\left(\frac{E}{T}\right)^{1/2}\frac{1}{s}
\end{equation}
Thus if the collision were to happen at the location $s_{bubble}$, then the center-of-mass energy would be
\begin{equation}
E_{cm}\sim m_p
\end{equation}
Since we are looking for interactions outside the location $s_{bubble}$, we see  that we should only consider center of mass energies that are below planck scale:
\be
E_{cm}<m_p
\label{beight}
\ee

Let us start by looking at the process $e^{-}+\gamma \rightarrow e^{-} \rightarrow e^{-}+graviton $. On dimensional grounds, we expect a cross section of the form 
\begin{equation}
\sigma\sim \frac{\alpha}{m^{2}_{p}}
\end{equation}
On the other hand the cross section of standard model processes has the form
\be
\sigma\sim \frac{1}{E_{cm}^2}
\end{equation}
Thus for $E_{cm}<m_p$ (eq. (\ref{beight})) , the above process involving a graviton gives a smaller probability of interaction than the standard model processes.

 Similarly, consider $e^{-}+graviton \rightarrow e^{-}\rightarrow e^{-}+graviton$. The cross section is
\begin{equation}
\sigma\sim \frac{1}{m^{4}_{p}}  E^{2}_{cm}
\end{equation}
Again, for $E_{cm}<m_p$, this is smaller than the standard model answer.

When $E_{cm}>m_p$, all these processes are expected to be superceded by micro black hole formation. But for these values of $E_{cm}$ the collision will happen inside the location $s_{bubble}$, as we have already noted in section \ref{secest2}. Thus we we do not find a modified firewall from the domain $E_{cm}>m_p$ either.

\subsection{A general relation for the modified firewall}

Finally, we recast the condition for a modified firewall in a convenient form. We recall  that for standard model processes,   the location where the integrated probability $P_{interact}$ became order one has the qualitative form given by (\ref{interaction distance})
\begin{equation}
s_{interact}\sim r_{h}e^{-\left( {r_hE}\right)}
\label{kone}
\end{equation}
A modified firewall exists if 
\be
s_{interact}\gg s_{bubble}
\label{condition2}
\ee
 where
\begin{equation}
s_{bubble}\sim \left(\frac{E}{T}\right)^{1/2}l_p\sim \sqrt{Er_h} \, l_p
\end{equation}
Thus the condition (\ref{condition2}) becomes
\be
r_{h}e^{-\left( {r_hE}\right)}\gg \sqrt{Er_h} \, l_p
\ee
which is
\be
r_h E \ll \h \log\left ( { r_h\over l_p^2 E}\right )
\label{condition3}
\ee
While the argument of the log can be large, the log itself is not, so let us set it as $\sim 1$. Then the condition (\ref{condition3}) becomes
\be
r_h\ll {1\over E}
\ee
That is, the radius of the black hole should be less than the wavelength of the infalling particle. For our case of an infalling electron, we can get a firewall behavior only if the radius of the hole is smaller than the Compton wavelength of the electron.

\section{Low energy scattering}\label{seclow}

Thus far we have considered the infall of particles which started at infinity with an energy $E\gg T$. This corresponds to $\lambda\ll r_h$; i.e., the wavelength of the infalling quantum is less than the radius of the hole.  In this energy domain we have seen that for macroscopic holes, the particle enters the bubble radius $s_{bubble}$ before it has a significant chance of interaction with any Hawking radiation quanta. Let us now consider the opposite domain
\be
E\lesssim T
\ee
To be in this domain we must consider massless infalling quanta; we take these quanta to be photons. Further,  the wavelength of the infalling quantum is now {\it larger} than the radius of the hole: $\lambda\gtrsim r_h$. 
The general analysis of \cite{flaw} (reproduced in section \ref{secest2}) says that in this case $s_{bubble} \lesssim l_p$. Since the fuzzball surface itself is at $s\sim l_p$, in this case there is no notion well defined `bubble' arising from the backreaction of the infalling quantum.  We will find that an infalling photon can  scatter off the electrons   present in the near-horizon Hawking radiation. But we will also find, using the analysis of section \ref{secgeodesic}, that unless this photon is scattered in a direction very close to the radially outwards direction, it cannot escape to infinity. 

\subsection{Low energy scattering cross section}\label{low e p}

Consider an electron of mass $m$ which is at rest. A photon of energy $\omega$ scatters off this electron. The scattering cross section is
\bea
\sigma&\sim& {\alpha^2\over m^2}, ~~~~\omega\lesssim m\nn
\sigma&\sim& {\alpha^2\over m \omega}, ~~~~ \omega\gtrsim m
\eea
Suppose we start with a photon that has energy $\omega\sim T$ at infinity. Then the energy of this infalling photon $\hat \omega(s)$ in the local orthonormal frame at any distance $s$ from the horizon will be of the same order as the Hawking temperature $\hat T(s)$ at that location
\be
\hat \omega(s)  \sim \hat T(s)
\ee
We would like to compute the scattering off electrons in the Hawking radiation. The outermost location where such electrons will exist in the Hawking radiation is  the location $s_m$ where the local temperature is order $m$
\be
\hat T \sim {1\over s_m} \sim m
\label{kthree}
\ee
When the infalling photon with energy $\omega\sim T$ reaches the location $s_m$, its local energy is 
\be
\hat \omega\sim \hat T \sim m
\label{ksix}
\ee
Thus the interaction cross section  is
\be
 \sigma \sim {\alpha^{2} \over m^2}
 \ee
 The probability for interaction $P$ is then given by
 \be
 {dP_{interact}\over ds}\sim \sigma \hat{n} 
 \ee
 The number density of electrons at this location is
 \be
 \hat{n}\sim {1\over s_m^3}
 \ee
To get an interaction at a distance $\sim s_m$ from the horizon, we consider the infalling quantum from the location $s_m$ to $s_m/2$. Then we get for the probability of interaction in this interval
\be
 P_{interact}\sim s_m {dP_{interact}\over ds} \sim  s_m \sigma \hat n \sim {\alpha^2\over m^2} {1\over s_m^2}
 \ee
 Using (\ref{kthree}), this is
 \be
 P_{interact}\sim \alpha^2
 \ee
 Thus we have a probability $P_{interact}\sim 10^{-4}$ for the infalling low energy photon to scatter off the Hawking radiation electrons at the outermost location where such an electron gas exists. This probability is not as small as the interaction probabilities we have encountered in computations of the above sections, and it does not decrease when we consider larger and larger holes. 
 
 \subsection{Probability for emergence}
 
 If we are to observe a photon which scatters in the above fashion, then this photon must emerge back to infinity, or at least to a region $r-r_h\gtrsim r_h$ away from the horizon. We have learnt in section \ref{secgeodesic} that if a particle starts near the horizon, then it will not emerge to infinity unless it starts out in a direction very close to the outward normal. Let us now see how small this solid angle for emergence is for our situation; i.e., for the case when the photon starts at the location $s_m\sim 1/m$.

The relation (\ref{finalc}) for the maximal angle of escape gives
\be
\sin\Psi \sim \Psi \sim  {s_m\over r_h}\sim {1\over m r_h}
\label{kfive}
\ee

The differential scattering cross section for scattering in backward direction $\theta=\pi$ is given in Appendix (\ref{P2 appendix}). We see that the cross section is of the same order in almost all directions including the backward scattering direction $\theta=\pi$. (The cross section becomes large in the forward direction $\theta=0$, but photons scattered in this forward direction will fall into the hole and not emerge to infinity.) Thus the probability of emergence is given by the solid angle corresponding to (\ref{kfive})
\be
P_{emergence} \sim {\Psi^2}\sim {1\over (m r_h)^2}
\ee

\subsection{Overall probability of backscattered photons}

We have sent in a photon with energy $\omega\sim T$ at infinity. This photon had a probability of scattering off the electron gas in the  radiation present near the fuzzball boundary, around the location $s_m\sim 1/m$ with a probability
\be
P_{interact}\sim \alpha^2\sim 10^{-4}
\ee
The scattered photon then had a probability of emergence to infinity given by 
\be
P_{emergence} \sim {1\over (m r_h)^2}
\ee
Thus the overall probability for scattering the photon back to infinity is
\be
P_{backscatter} \sim P_{interact}  \, P_{emergence}\sim {\alpha^2 \over (m r_h)^2}
\label{pback}
\ee
A more detailed calculation of $P_{backscatter}$ can be found in Appendix {\ref{P2 appendix}}. For reasons that we will mention below, let us assume that for now $\omega/T>1$. Using the result (\ref{finalb}), we compute  $P_{backscatter}$ as a function of $\omega/T$ for a solar mass hole. These values of $P_{backscatter}$  are shown in fig.\ref{pc}, in the part of the graph for $\omega/T>1$.   We see that 
$P_{backscatter}\ll 1 $ everywhere in this domain.

\subsection{Wavelengths $\lambda \gg r_h$}

Let us also ask for the scattering probability when the wavelength of the infalling photons is much {\it larger} than the radius $r_h$ of the black hole. Consider a spherical wave incident on the hole from infinity. In the limit $\lambda \gg r_h$,  there is a small probability $P_{barrier}$ for an incoming spherical wave to penetrate to the near horizon region; the rest of the wave gets reflected out to infinity by the barrier in the effective potential present in the near horizon region \cite{unruh}. For a scalar field we have $P_{barrier} \sim ( r_h/\lambda)^2\sim (\omega/T)^2$. For a vector field like the photon, we have $P_{barrier} \sim ( r_h/\lambda)^4\sim (\omega/T)^4$. 

\begin{figure}
  \centering
    \centering
    \includegraphics[width=4in]{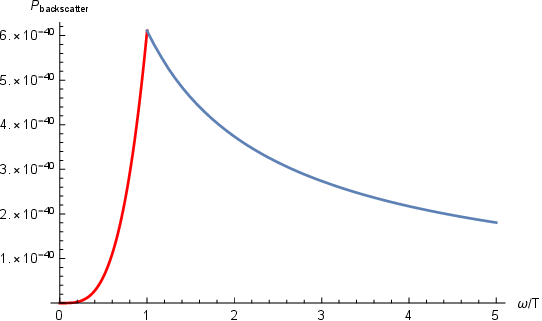}
    \caption{ $P_{backscatter}$ for different values of  $\omega/T$ for a solar mass hole. The red line (the left part of the  curve) uses the greybody factor expression for $\omega/T\ll 1$, while the   blue line (right side of the curve) sets the greybody factor to unity, which is appropriate for $\omega/T \gg 1$. }
    \label{pc}
\end{figure}

Since our infalling photon is spin 1,  for the case $\lambda \gg r_h$ we must multiply the probability $P_{backscatter}$ in eq.(\ref{pback}) by $(\omega/T)^4$. In fig.\ref{pc} we plot this product for the domain $\omega/T<1$. The full plot should be smooth everywhere. The approximation  in fig.\ref{pc}  has a cusp at $\omega/T=1$ because we have not done a precise computation of the greybody factor $P_{barrier}$; instead we have patched together the rough estimates  for the domain $\omega/T<1$ and the domain $\omega/T>1$.\footnote{One might wonder if there should be another factor $P_{barrier}\sim (\omega/T)^4$ for the scattered wave to {\it escape} back to infinity. But in fact there will not be such a factor.  After the $\lambda \gg r_h$ wave scatters off the electrons in the radiation, the scattered wave will have an energy $\omega\sim 1/m$. This is not small; in fact it is of order the local Hawking temperature. The greybody factors for Hawking quanta are order unity, so we do not have  any further suppression factor $(\omega/T)^4$ coming from the escape of the scattered photon to infinity.}

Overall, we note that $P_{backscatter}\ll 1$ everywhere for our present case of low energy scattering.  The peak is around $\omega/T\sim 1$, where we have
\be
P_{backscatter} \sim 10^{-39}
\label{kseven}
\ee

\subsection{Observing these backscattered photons}

 From (\ref{ksix}) we see that at the location of scattering, in our orthonormal frame, the energies of the photon and the electron are comparable
 \be
 \hat \omega\sim \hat T \sim m
 \ee
 In such a situation, the energy of the scattered photon will also be
 \be
 \hat{\omega}\sim \hat T
 \ee
 If this photon emerges to infinity, its energy at infinity will then be
 \be
 \omega\sim T
 \ee
 That is, the energy of the backscattered photon will be of the same order as the Hawking temperature.

 But the hole is emitting photons  with energy $\omega\sim T$, by virtue of its temperature,  even in the absence of any probe photons that have been sent in. There are $\sim 1$ such photons per mode. That is, if we look at infinity for photons of wavelength $\lambda \sim r_h$, in a box of radial extent $\Delta r \sim r_h$, then at any given time we expect to find one such Hawking photon in that box.
 
 Thus if we are to send in additional $\omega\sim T$ photons and observe them after backscattering, then we have to be able to distinguish these backscattered photons from the photons of Hawking radiation. So we need $\gg 1$ photons to be present in the same radial box of width $\Delta r\sim r_h$ that we mentioned above. Given the backscatter probability (\ref{kseven}), we will need to send in a pulse of more than $N$ photons where
 \be
 N\sim 10^{39}
 \ee
 in order to detect the backscattering of the photons off the electron gas.

 \subsection{Backscattering off primordial holes}
 
 Black holes formed by stellar collapse are order solar mass. But if primordial holes form, then they may have smaller sizes. If these holes formed at the big bang and had a mass $M\lesssim 10^{11}\, kg$ then they would have evaporated by now. Thus consider black holes that are slightly bigger; say $M\sim 10^{12}\, kg$. The horizon radius for such holes is
 \be
 r_h\sim 10^{-12}\, m \sim \lambda_c
 \label{lambdacb}
 \ee
 where $\lambda_c\sim 1/m$ is the Compton wavelength of the electron. Then we find that
 \be
P_{emergence} \sim {1\over (m r_h)^2}\sim 1
\ee
Thus
\be
P_{backscatter} \sim {\alpha^2 \over (m r_h)^2}\sim \alpha^2 
\label{pb1}
\ee
This probability is not small. But the primordial black hole has a small radius, and so the overall cross section for scattering anything off the hole is small. First consider scattering off the region $r\sim r_h$, which is what we have considered in our discussions above.  Suppose we were to probe the hole with photons of wavelength $\lambda_c$. Taking the backscattering probability (\ref{pb1}) into account, the effective scattering cross section from such a hole will be
\be
\sigma_{eff}\sim r_h^2 P_{backscatter} \sim 10^{-4}\lambda_c^2 
\label{scatteringba}
\ee
This is a very small cross section, and we are unlikely to see these black holes by such a  scattering. Note that (\ref{scatteringba}) is the same as the Compton scattering cross section of a free electron in flat space; this is the case because when $r_h\sim \lambda_c$, the electron we scatter off is not close to the horizon but at a distance $\sim r_c$ from the horizon.

Now we note that since these black holes have $T\sim m$, they emit electrons as their radiation. These emitted electrons  form a cloud around the hole, and we can try to scatter off this cloud. By the general properties of Hawking radiation, we expect $\sim 1$ emitted electron in each annular region  $n r_h<r<(n+1)r_h$, where $n=1,2,\dots$. 

Suppose we probe the hole with photons of wavelength $\lambda \gg r_h$. Then we probe the above cloud of emitted electrons out to a radius $\sim \lambda$. Thus the ball of radius $\lambda$ around the hole   contains
\be
N\sim {\lambda\over r_h}\sim {\lambda\over \lambda_c}
\ee
electrons, where we have used (\ref{lambdacb}). The Compton scattering cross section of each such electron is $\sigma_{e\gamma}\sim \alpha^2/m^2$. Thus the total cross section of the electrons in this ball is
\be
\sigma_{total}\sim N \sigma_{e\gamma}\sim {\lambda\over \lambda_c}{\alpha^2\over m^2}\sim \alpha^2{\lambda\over m}
\ee
Since the incoming wave had a transverse area $\gtrsim \lambda^2$, we see that the probability of scattering is
\be
P_{scattering} \sim {\alpha^2{\lambda\over m}\over \lambda^2} \sim {\alpha^2\over \lambda m}\sim \alpha^2{\lambda_c\over \lambda}
\ee
For $\lambda \gg \lambda_c$ this is much less than unity, so it is not easy to scatter off the cloud of electrons radiated by the primordial hole.  

One might think that if there is a large number of such black holes per unit volume then the scattering could be seen, but note that each black hole is very heavy, and the number density of such holes is limited by the fact that they should not over-close the universe.

\subsubsection{Photon-Photon Scattering}
 
In the above discussion we have considered the infall of a low energy photon and its scattering off the electron gas present in the radiation near the fuzzball surface. Such electrons are present only where the local temperature becomes of order the electron mass; i.e. for distances less that 
\be
s\sim {1\over m}
\ee
On the other hand photons are massless, and so are present in the radiation at much larger values of $s$. Thus one might wonder if it is relevant to consider photon-photon scattering, where the infalling photon scatters off a photon in the radiation bath. The cross section for this scattering is small since it is a 1-loop process, but this might be compensated by the fact that photons are present in the radiation bath at larger $s$, and we have seen that it is easier for scattered quanta to escape from infinity if they start at a larger $s$. 

But a simple estimate shows that photon-photon scattering is a much smaller effect that the photon-electron scattering that we have already considered above. Consider the infalling photon to have energy  at infinity $\omega\sim T$. Then we have from (\ref{COM})
\be
E_{cm}\sim {1\over s}
\ee
First consider the region $s\lesssim 1/m$. In this region 
$E_{cm}\gtrsim m$. For such energies we have 
\be
\sigma_{\gamma\gamma}\sim {\alpha^4\over E_{cm}^2}
\ee
so this is smaller than the Compton cross section by a factor $\alpha^2$. 

Now consider the region $s\gtrsim 1/m$; this is the region where photons exist in the radiation, but  electrons do not. Now $E_{cm}<m$, and the photon-photon scattering cross section is  
\be
\sigma_{\gamma\gamma}\sim {\alpha^4E_{cm}^6\over m^8}
\ee
Using this we can compute the probability of backscattering as before
\be
{dP_{backscatter}\over ds}\sim \sigma_{\gamma\gamma}(s) \hat n(s) P_{emergence}(s) \sim \sigma_{\gamma\gamma}(s) \hat n(s) {s^2\over (GM)^2} \sim {\alpha^4\over m^8 (GM)^2} {1\over s^7}
\ee
where we have used  $\hat n(s) \sim {1/s^3}$ and the maximal angle of escape (\ref{finalc}) for $P_{emergence}$.

We see that the scattering probability falls off very rapidly with $s$. Thus the largest contributions to scattering in the region $s\gtrsim 1/m$ will come from $s\sim 1/m$. But in the region $s\sim 1/m$ where we have already seen that the scattering off electrons is larger than the scattering off photons. 

We can restate the above conclusion by noting that the box diagram for photon-photon scattering involves an electron loop, and the cross section will be small unless we reach energies which are of order the electron mass. But once we reach energies of order the electron mass, we can scatter off electrons in the radiation with a larger probability since Compton scattering is $\sim \alpha^2$ while photon-photon scattering is $\sim \alpha^4$.

\section{Discussion}\label{secdiscussion}

The traditional picture of a black hole has a horizon, and the spacetime around this horizon is in the vacuum state. Such a picture however leads to the information paradox. In string theory, it appears that the situation is very different: instead of a spacetime with horizon, we get a horizon sized `fuzzball' which radiates unitarily from its surface just like any other warm body.

One may then ask if it is possible to observe the difference between the traditional hole and a fuzzball. In this paper we have argued that observing this difference will be very difficult, mainly because the differences are located very close to the radius $r_h=2GM$. 
We have argued that the fuzzball boundary is likely to be only order  planck length outside the horizon radius $r=r_h$; we called such a model a `tight fuzzball' as opposed to a `diffuse fuzzball'. 

With a tight fuzzball, we still have the option of interacting directly with the fuzzball surface, or with the radiation emerging from the surface. The temperature of the radiation bath drops quickly with the distance $s$ from the location $r=r_h$, so even though this radiation is outside the fuzzball boundary, interactions with this radiation are probable only close to the fuzzball surface. 

A further issue is the backreaction of the infalling quantum on the fuzzball. We have noted that if causality is to be preserved, the fuzzball surface must bulge out to absorb an incoming quantum once this quantum reaches a location $s=s_{bubble}$. Thus any interactions with the radiation bath must occur outside this radius $s_{bubble}$; otherwise we are interacting with the fuzzball surface itself. 

With this setup, we made the following observations:

\b

(a) If we have a tight fuzzball, i.e., one whose surface is order planck length outside the  $r_h=2GM$, then it is very hard to get direct observational evidence for such a surface. There is a combination of reasons for this difficulty:

\b

\quad (i)  One might think of sending a light ray that grazes the fuzzball surface and comes out to infinity; this can indicate the precise location of the surface. But the near horizon behavior of trajectories in the Schwarzschild metric does not allow such grazing rays to exist. If a ray comes in from any  location with $r-r_h\gtrsim r_h$, then it reaches points close to $r_h$ in a near-radial direction (eq. (\ref{nearradial})). 

\b

\quad (ii) Once a quantum does hit the fuzzball surface, then it has a high probability of converting its energy to excitations of the fuzzball as opposed to reflecting off the fuzzball surface (eq.\ref{preflect}). The reason for this is the very large density of states of the fuzzball (eq. (\ref{bthree})), which is a basic feature of the black hole: its very large entropy. Thus the fuzzball surface is highly absorptive, and in this sense mimics the horizon of a traditional hole to high accuracy. 

\b

\quad (iii) If any part of the incident quantum does manage to reflect off the fuzzball surface, then it typically not be able to escape out to the region $r-r_h\gtrsim r_h$; the only rays that escape from near the horizon are those that start off vary close to the outward radial direction (eq. (\ref{finalc2})).

\b

(b) Instead of scattering a quantum off the fuzzball surface, one might try to scatter it off the {\it radiation} emerging from the surface. The firewall argument had envisaged an infalling object getting `burnt' by high energy radiation emanating from the such a surface, with the expectation that the encountered radiation would hotter and hotter as the infalling object approached $r=r_h$. But a flaw in this reasoning was pointed out in \cite{flaw}. Even in classical gravity, we find that the horizon expands out teleologically to engulf an infalling particle {\it before} this  particle reaches $r=r_h$. An object like a fuzzball must have its surface always outside this classical horizon; otherwise information will not be able to emerge from the hole without violating causality \cite{causality1,causality2}. The entropy-enhanced tunneling mechanism in the fuzzball paradigm provides a mechanism for the fuzzball surface to extend out in the required manner to engulf the infalling quantum.

Once the fuzzball surface reaches the infalling quantum, then we have the same difficulty with reflecting the quantum that we noted  above: the fuzzball has a high density of states so it absorbs any energy falling on it with a very high probability. So we should ask if the infalling quantum can have a significant interaction with the radiation {\it before} it gets engulfed by the fuzzball surface. The fuzzball surface expands out to a location $s_{bubble}$ given by eq.(\ref{ssbubble}). Thus we need to find the probability $P_{interact}$ for the quantum to interact with radiation before the quantum reaches the location $s_{bubble}$; if we do find a probability $P_{interact}\sim 1$ for such an interaction then we say that we have a `modified firewall'.

It had already been noted in \cite{flaw} that in a theory with no large dimensionless numbers, the dominant interaction contributing to $P_{interact}$ was given by the gravitational process of mini-black hole formation, and that this interaction did not give a modified firewall for macroscopic holes. This still left open the question of whether a modified firewall could be found for astrophysical holes  in the standard model, which does have a large dimensionless number arising from the hierarchy of coupling constants. But for the processes we studied, we found no evidence of a modified firewall.

\b

(c) The electron-photon interaction cross section decreases with energy. So we can ask: what happens if we use wavelengths that are larger than the  radius of the hole $\lambda \gtrsim r_h$? This domain was not covered in case (b) above, since  there we envisaged a quantum that could `fall' into the hole. 

For $\lambda\gtrsim r_h$, we find that an infalling photon can scatter off the radiation before reaching the fuzzball surface. But the probability for the scattered photon to be able to escape the near horizon region remains small, since this scattering happens very close to the horizon. So it appears hard to detect the fuzzball this way also.

\b

(d) All the above considerations (a)-(c) were carried out in the domain where the infalling quantum was much  lighter than the mass of the black hole. It is possible that the situation changes when the energy of the infalling perturbation approaches the mass of the hole. In this case the new surface of the fuzzball will form far outside the location $r=r_h$, and it is easier for rays to escape from such a location.

 It was argued in \cite{causality1,causality2} that spacetime near the horizon should be modeled like a lake: infalling quanta are like waves on this lake, and $r=r_h$ is like the boundary of the lake. Quantum gravity effects like tunneling into fuzzballs happen when the wave amplitude is large enough to touch the bottom of the lake. When a wave has an amplitude large enough to touch the bottom of the lake, then a part of this wave can be  reflected back; the exact details of how {\it much} is reflected back depends on the depth profile of the water. 
 
  It is conceivable that a similar reflection from happens for large amplitude gravitational  waves from the location where they re about to form a new horizon because of their backreaction on the geometry.  In that case we would indeed get a signal of quantum gravity effects from gravitational waves. To learn whether any such signal should actually arise, we would need a more detailed study of the nonlinear dynamics of fuzzballs. 
 
 But for the most part the physics of gravitational waves can be carried out in the domain of perturbative gravity. In this domain the above considerations (a)-(c) apply, and these considerations do not imply any novel signals from gravitational waves.

 \b

Let us now  comment on the significance of the fact that it is difficult to distinguish the fuzzball from a black hole by simple observations. It was argued in \cite{fuzzcomp} that this difficulty signals a general `correspondence principle'. This principle says that for external observations involving  $E\gg T$ quanta , the fuzzball can be replaced by the traditional hole to good accuracy.  

Finally, we note some of the attempts to find measurable signals of deviations from the traditional hole. In \cite{cardoso2,afshordi} it was noted that reflections off horizon would lead to echoes in the radiation.  Some features of the data already obtained could be interpreted as evidence for such echoes. In \cite{hartle}, it was noted that the formation of fuzzballs could give unique observable signatures. It was also noted that the spread of the wavefunction over the space of fuzzballs could give a novel kind of gravitational wave burst. In \cite{cardoso} a detailed analysis was given of different structures near the horizon, and what these structures might imply for the observability of quantum gravity effects. In \cite{pen} it was noted that it may be possible to observe the interference of pulsar radiation with the radiation emitted by a black hole, and thereby see signatures of quantum gravity.  

To conclude, we have seen that it is difficult to get direct observational evidence of a tight fuzzball, when using probes with energy $E$ much less than the mass of the hole. If observations do indicate departures from novel quantum gravity effects, then these would be evidence for diffuse fuzzballs. Also, gravitational waves emitted by black hole mergers involve energies that are comparable to the mass of the hole, and in this situation we may have novel effects of the kind discussed in point (d) above. 

 \b
 
 \b

 \section*{Acknowledgements}

We would like to thank Niayesh Afshordi,  Vitor Cardoso,  Stuart Raby,  David Turton and Hong Zhang for helpful discussions.  This work is supported in part by DOE grant de-sc0011726.

\pagebreak

\appendix

\section{Computing the interaction probability for an infalling electron}\label{P appendix}

Here we derive an expression for the interaction probability of an electron with gas of photons. We start with the situation where the electron is at rest, and later transform to a frame where the electron is moving.

Consider the differential cross-section for Compton scattering where  an electron is at rest and it is being bombarded by a gas of photons:
\bea
{d \sigma \over d \cos\theta'_{rest}} &=& {\pi \alpha^2\over m^2}{ 1 \over \big( 1 + {\omega^{rest} \over m}(1-\cos\theta'_{rest})\big)^2}\cr
&&\qquad \Bigg[ { 1 \over 1 + {\omega^{rest} \over m}(1-\cos\theta'_{rest})} + 1 + {\omega^{rest} \over m}(1-\cos\theta'_{rest})-(1-\cos^2\theta'_{rest}) \Bigg]\nn
\label{differential cross section}
\eea
Here $\omega^{rest}$ is the photon energy before interaction,  and $\theta'_{rest}$ is the angle through which the photon scatters. Integrating (\ref{differential cross section}) over $\cos\theta'_{rest}$ from $-1$ to $1$ yields the full scattering cross section in the rest frame of electron:
\bea
\sigma = 2{\pi\alpha^2\over m^2}\Big\{\frac{1+x}{x^{3}}\Big[\frac{2x(1+x)}{1+2x}-\ln(1+2x)\Big]+\frac{1}{2x}\ln(1+2x)-\frac{1+3x}{(1+2x)^{2}}\Big\}
\label{cross section}
\eea
where $x={\omega^{rest}\over m}$. Note that the total cross section does not depend on the initial photon angle because of rotational symmetry;  in (\ref{differential cross section}) we have already averaged over initial spins and summed over final spins.

Next we compute the expression for the interaction probability. Let $n_{rest}$ be the photon number density within an energy bin between $\omega^{rest}$ and $\omega^{rest} + d\omega^{rest}$ and a solid angle bin between $\Omega^{rest}$ and $\Omega^{rest} + d\Omega^{rest}$, where we are still in the rest frame of the electron.  Consider the interaction probability $P$ in a time $dt_{rest}$. Note that the photons move with the speed of light, so  they move a distance $dl=dt_{rest}$ in this time. We then have
\bea
P = \int \sigma n_{rest} dl=\int \sigma n_{rest} dt_{rest}
\label{probability}
\eea

In our physical problem, the electron is not at rest but is moving towards the black hole. To get the scattering probability in this case we will perform an appropriate Lorentz transformation.

Consider the frame where the electron is moving along the  $-z$ direction. We call this frame the lab frame. Let the electron have a velocity  $-\beta$, so that with our conventions, $\beta>0$. Let this electron  collide with a photon coming in at  an initial angle, $\theta_{lab}$; this angle is measured from the positive $z$ direction.

To get the scattering probability, we boost from the lab frame back into into the rest frame where the quantities in (\ref{probability}) are defined. This gives
\bea
\begin{pmatrix}
\omega^{rest}\\
k^{rest,\hat{x}}\\
k^{rest,\hat{y}}\\
k^{rest,\hat{z}}
\end{pmatrix}
&=&
\begin{pmatrix}
\gamma & 0 & 0 & \gamma\beta\\
0 & 1 & 0 & 0\\
0 & 0 & 1 & 0\\
\gamma\beta & 0 & 0 & \gamma\\
\end{pmatrix}
\begin{pmatrix}
\omega^{lab}\\
k^{lab,\hat{x}}\\
k^{lab,\hat{y}}\\
k^{lab,\hat{z}}
\end{pmatrix}
\eea
yielding
\bea
\omega^{rest} = \gamma \omega^{lab} (1+\beta \cos\theta^{lab}_{ph})
\eea
Similarly, we can apply this same Lorentz transformation to the number current 4-vector of photon $(n_{lab}, n_{lab}\vec v)$. This yields
\bea
n_{rest} &=& \gamma(1+\beta \cos\theta_{lab})n_{lab}
\label{boosted1}
\eea
Finally, we apply the Lorentz transformation to the infinitesimal displacement 4-vector of the electron $(dt_{lab}, 0, 0, -ds_{lab})$. This gives
\bea
dt_{rest}&=&\gamma dt_{lab}-\gamma\beta ds_{lab}=\gamma ds_{lab}(\frac{1}{\beta}-\beta )\nn
&=&{1\over\gamma\beta} ds_{lab}
\label{boosted quantities}
\eea
where $ds_{lab}$ is the distance that electron moves along $-z$ direction in lab frame.
Inserting (\ref{boosted1}) and (\ref{boosted quantities}) into (\ref{probability}) yields the expression
\bea
P = \int {( 1 + \beta\cos\theta_{lab})\over \beta}\sigma n_{lab} ds_{lab}
\label{P 2}
\eea
Here, $\sigma$ is given by (\ref{cross section}) with $x\equiv{\omega^{rest}\over m} = \gamma {\omega^{lab}\over m}(1+\beta\cos\theta_{lab})$. 

We now remove the label `lab' from each variable since we'll only be working in the lab frame for the remainder of the section. Thus all the quantities should be understood in the lab frame.

For a thermal photon gas at temperature $T$, the photon number density within an energy bin $\omega$ and $\omega + d\omega$ and a solid angle bin $\Omega$ and $\Omega + d\Omega$, $n$, is given by the expression
\bea
n = \frac{1}{4\pi^{3}}\frac{\omega^{2}}{e^{\omega/T}-1}d \omega d \Omega
\label{number density}
\eea
Then (\ref{P 2}) gives
\bea
P &=&  \frac{\alpha^2}{\pi } \int   {1\over \gamma^2\beta^2}\frac{1}{e^{\omega/T}-1} ds d \omega\cr
&&~~\times \int_{{\gamma \omega\over m}(1 - \beta)}^{{\gamma \omega\over m}(1+\beta)}    \left(\frac{1+x}{x^{2}}\Big[\frac{2x(1+x)}{1+2x}-\ln(1+2x)\Big]+\frac{1}{2}\ln(1+2x)-\frac{x(1+3x)}{(1+2x)^{2}}\right)   dx  \nn
\eea
Performing the integral over $x$ yields
\bea
P &=&  \frac{\alpha^2}{\pi } \int   {1\over \gamma^2\beta^2}\frac{1}{e^{\omega/T}-1}  \left[F\left(\frac{\gamma \omega (1+\beta)}{m}\right)-F\left(\frac{\gamma \omega (1-\beta)}{m}\right)\right]ds d \omega \nn
\label{P rest 3}
\eea
where 
\bea
F(y)&=&\frac{1}{8}\bigg[-2-2y+\frac{1}{1+2y}+(18+8\frac{1}{y}+4y)\log(1+2y)+8 \text{Polylog}(2,-2y)\bigg]\nn
\eea
with the limits
\bea\label{Flow}
F_{0}(y)&=&\frac{15}{8}+\frac{2y^{2}}{3}+O(y^{3})~~~y\rightarrow 0
\eea
\bea\label{Fhigh}
F_{\infty}(y)&=&\frac{1}{4}(-y+2y\log(2y))~~~~y\rightarrow \infty
\eea
Here we've computed an expression for the full interaction probability of an infalling electron interacting with a thermal gas of photons, in the frame where this gas is at rest.

\section{Computing the interaction probability for an infalling photon}\label{P2 appendix}

Consider a photon that is falling in radially near the fuzzball surface. In the region where the temperature is high enough, the radiation contains electrons and positrons in the thermal bath. We are interested in the probability that the infalling photon scatters off this electron-positron bath. 

\subsection{The kinematics}

We work in a local orthonormal frame  oriented along the Schwarzschild coordinates. We orient the $z$ direction along the outward radial direction. Thus the infalling photon has 4-momentum
\be
k=(\omega, 0, 0, -\omega)
\ee
Let the photon collide with an  electron with mass $m$ in the thermal bath. This electron has an energy $E$ and its momentum direction is described in polar coordinates by  $(\theta_e, \phi_e)$ defined by our choice of $z$ axis. Thus the incoming electron has 4-momentum
\be
p=(E,  p\sin\theta_e\cos\phi_e, p\sin\theta_e \sin\phi_e, p\cos\theta_e)
\ee
where
\be
E=\sqrt{p^2+m^2}
\ee
Let the scattered electron have energy $E'$ and momentum direction $(\theta'_e, \phi'_e)$. Let the scattered photon have energy $\omega'$ and direction $(\theta'_{ph}, \phi'_{ph})$. Thus the outgoing photon has 4-momentum
\be
k'=(\omega', \omega'\sin\theta'_{ph}\cos\phi'_{ph}, \omega'\sin\theta'_{ph}\sin\phi'_{ph}, \omega'\cos\theta'_{ph})
\ee

\subsection{Scattering rate}

In electron-photon scattering, the number of scattering events per unit volume per unit time is given by
\be
N_{scatter}=\int |M|^2 (2\pi)^4 \delta^4(p+k-p'-k') {d^3 p'\over (2\pi)^3 2E'}{d^3 k'\over (2\pi)^3 2\omega'}
\label{bfive}
\ee
To use this expression for our situation, we note the following factors:

\b

(i) The  expression (\ref{bfive}) assumes $2\omega$ photons per unit volume. Let us work in a volume $V$. This would contain $2\omega V$ photons. We on the other hand have a single infalling photon. Thus we should multiply $N_{scatter}$ by the factor
\be
{1\over 2\omega V}
\ee

\b

(ii) The  expression (\ref{bfive}) assumes $2E$ electrons per unit volume. We on the other hand have an electron density given by the Fermi-Dirac distribution
\be
n(E, \Omega)={4\over (2\pi)^3} {E\sqrt{E^2-m^2} \, dE d(\cos \theta_e) d\phi_e\over e^{E\over T}+1}
\ee
where the factor of $4$ corresponds to 2 spins of electrons and 2 spins of positrons and we have written the phase space measure as
\be
{d^3 p\over (2\pi)^3}= {|p|^2 d|p|d\Omega_e\over (2\pi)^3}= {E|p| dEd\Omega_e\over (2\pi)^3}
={E\sqrt{E^2-m^2}dEd\Omega_e\over (2\pi)^3}
\ee
Thus we should multiply $N_{scatter}$ by the factor
\be
\left ( {1\over 2E} \right )  {4\over (2\pi)^3} {E\sqrt{E^2-m^2} \, dE d(\cos \theta_e) d\phi_e\over e^{E\over T}+1}
\ee

 \b
 
 (iii) We must multiply (\ref{bfive}) by the spacetime volume. We have taken the spatial volume to be $V$. We are looking at the interaction probability $dP_{interact}$ in the time the photon moves through a distance $ds$. This corresponds to a time $dt=ds$. Thus we should multiply (\ref{bfive}) by 
 \be
 Vds
 \ee
 
 \b
 
 With these additional  factors (i)-(iii), the number of scattering events becomes the scattering probability $dP_{interact}$. We find
\bea
&&{dP_{interact}\over ds}=\nn
 &&\int\!|M|^2 (2\pi)^4\delta^4(p+k-p'-k'){d^3 p'\over (2\pi)^3 2E'}{d^3 k'\over (2\pi)^3 2\omega'}{1\over (4 E\omega) }{4\over (2\pi)^3} {E\sqrt{E^2-m^2} \, dE d(\cos \theta_e) d\phi_e\over e^{E\over T}+1}\nn
\label{eqfull}
\eea

\subsection{Performing the computation}

We sum over the final spins and average over initial spins. This gives
\bea
 |M|^2 \!\r {1\over 4}\! \sum_{spins} \!|M_{spin}|^2\!={2\alpha} \left [ {p\cdot k'\over p\cdot k} \!+\!  {p\cdot k\over p\cdot k'}\!+\!2m^2 \left ( {1\over p\cdot k}\!-\!{1\over p\cdot k'} \right ) \!+\! m^4 \left ( {1\over p\cdot k}\!-\!{1\over p\cdot k'}\right )^2 \right ]\nn
\eea
We find
\bea
p\cdot k &=&\omega ( E+\sqrt{E^2-m^2}\cos\theta_e)\cr 
p\cdot k'&=&\omega' [ E- \sqrt{E^2-m^2}\cos\theta_e\cos\theta'_{ph}- \sqrt{E^2-m^2}\sin\theta_e\sin\theta'_{ph}\cos ( \phi_e-\phi'_{ph})]\nn
\eea

The momentum delta function can be easily solved 
\be
\int d^3  p' \delta^3 ( \vec k +\vec p-\vec k'-\vec p') =1
\ee
This sets $\vec p'= \vec k+\vec p-\vec k'$, so that $E'\sqrt{|p'|^2+m^2}$ is no longer independent of $|k'|=\omega'$. We write $\int d^3 k'= \int w'^2 d\omega' d\Omega'_{ph}$. The
energy delta function gives a Jacobian
\be
\int d\omega' \delta (E'+\omega' - E-\omega) = [1+{dE'\over d\omega'}]^{-1}
\ee
where
\be
{dE'\over d\omega'}= {[\omega'+ \omega  \cos\theta'_{ph} - \sqrt{E^2-m^2} [\cos\theta_e\cos\theta'_{ph}+ \sin\theta_e\sin\theta'_{ph}\cos ( \phi_e-\phi'_{ph})]\over E'}
\ee
Finally, the energy-momentum conservation give the relations
\bea
\omega'&=&{ \omega \sqrt{E^2-m^2} \cos\theta_e +\omega E\over (\omega+E)+ \omega \cos\theta'_{ph} -\sqrt{E^2-m^2} [\cos\theta_e\cos\theta'_{ph}+ \sin\theta_e\sin\theta'_{ph}\cos ( \phi_e-\phi'_{ph}))] }
\nn
E'&=&E+\omega-\omega'
\eea
Putting all these relations in (\ref{eqfull}) gives ${dP_{interact}\over ds}$.

\begin{figure}[h]
  \centering
    \centering
    \includegraphics[width=4in]{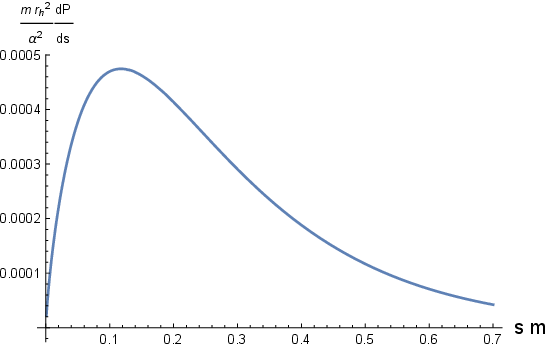}
    \caption{$dP_{backscatter}/ds$ as a function of the location $s$.  The $x$-axis gives $s$ in units of the Compton wavelength of the electron $1/m$. The $y$-axis gives  $dP_{backscatter}/ds$ in units of $\frac{\alpha^2}{mr_{h}^2}$. }
    \label{dpdsB}
\end{figure}

\subsection{Backward scattering}

The expression we get from the above computation is complicated, but we can simplify it by noting that we are interested in the case where the scattered photon is emitted very close to the outward normal direction of the black hole; only in this case can it escape the near horizon region. Thus we set
\be
\theta'_{ph}=0
\ee
Further, the temperature is
\be
T\approx {1\over 2\pi s}
\ee
The solid angle $d\Omega'_{ph}$ along which the photon can emerge to infinity is given by (\ref{finalc}); thus we have
\be
d\Omega'_{ph}={27\over 16} { s^2\over r_h^2}
\ee
With these substitutions, we find
\bea
&&\hskip -.2 truein {dP_{backscatter}\over ds}={\alpha^2\over 2 \pi^2}\times  \nn
&& \hskip -.2 truein \int\!\Bigg [2+  {4 m^4\over (E^2\sin^2\theta_e+m^2 \cos^2\theta_e)^2}+ {2 E \omega - 4 m^2\over E^2\sin^2\theta_e+m^2 \cos^2\theta_e} 
- { 2 \omega (E+2\omega)\over (E+2\omega)^2-(E^2-m^2)\cos^2\theta_e}
\nn
&& + { 8 \sqrt{E^2-m^2} \, \omega^2 (E+\omega)\cos\theta_e\over E^2(E+2\omega)^2+(E^2-m^2)\cos^2\theta_e[(E^2-m^2)\cos^2\theta_e - 2 ( E^2+2E\omega + 2\omega^2)]} ~ \Bigg ]
\nn
&& \times { E+ \sqrt{E^2-m^2} \cos\theta_e\over [E+2\omega-\sqrt{E^2-m^2}\cos\theta_e ]^2}
\times  {27\over 16} {s^2\over r_h^2}
\times
 {\sqrt{E^2-m^2} \, dE d(\cos \theta_e) \over e^{2\pi s E }+1}\nn
 \label{finalb}
\eea

To illustrate this result, we plot $dP_{backscatter}/ds$ in Fig.\,\ref{dpdsB}. We consider energies for the infalling photon that are of order   $\omega\sim T$.
We find that $dP_{backscatter}/ds$ has a peak in the region where $s$ is of order $1/m$. 

Let us now check that the full computation above reproduces the estimates that we have used in section \ref{seclow}. The height of the graph of $dP_{backscatter}/ds$ is order  $\frac{\alpha^2}{m r_{h}^2}$ and the width of the peak is of order  $\Delta s \sim 1/m$. We can estimate $P_{backscatter}\sim {dP_{backscatter}\over ds} \times \Delta s$:
\begin{equation}
P_{backscatter}\sim \frac{\alpha^2}{m r_{h}^2} \frac{1}{m}\sim  {\alpha^2 \over (m r_h)^2}
\end{equation}
which is the same as the estimate  (\ref{pback}).

\end{document}